\documentclass[12pt]{aastex}
\usepackage{latexsym,graphicx,natbib}

\newcommand{\beq}{\begin{equation}}
\newcommand{\eeq}{\end{equation}}

\begin{document}
\bibliographystyle{icarus}

\title{Photometric Survey of the Irregular Satellites}

\medskip

\author{Tommy~Grav\altaffilmark{1,2,3}}
\affil{\footnotesize \it Institute of Theoretical Astrophysics, University in Oslo,\\
	Postbox 1029 Blindern, 0359 Oslo, Norway (tommy.grav@astro.uio.no) \\
	 \& \\
	Harvard-Smithsonian Center for Astrophysics, \\ 
	MS51, 60 Garden Street, Cambridge MA 02138 (tgrav@cfa.harvard.edu)}
\author{Matthew~J.~Holman\altaffilmark{2,3}}
\affil{\footnotesize \it Harvard-Smithsonian Center for Astrophysics, \\
MS51, 60 Garden Street, Cambridge, MA 02138 (mholman@cfa.harvard.edu)}
\author{Brett~J.~Gladman}
\affil{\footnotesize \it Dept. of Physics and Astronomy, University of British Columbia\\
6224 Agricultural Road, Vancouver, B.C., V6T 1Z1, Canada (gladman@astro.ubc.ca)}
\author{Kaare~Aksnes}
\affil{\footnotesize \it Institute of Theoretical Astrophysics, University in Oslo, \\
Postbox 1029 Blindern, 0359 Oslo, Norway (kaare.aksnes@astro.uio.no)}

\altaffiltext{1}{Visiting Astronomer, Nordic Optical Telescope}
\altaffiltext{2}{Visiting Astronomer, MMT Observatory}
\altaffiltext{3}{Visiting Astronomer, Magellan Observatory}

\medskip

\date{\rule{0mm}{0mm}}

\newpage

{\bf Proposed Running Head}: Photometric Survey of Irregular Satellites

\vspace{1cm}
{\bf Editorial Correspondence to}: \\
Tommy Grav \\
MS \#51, 60 Garden Street \\
Cambridge, MA 02138 \\
USA \\
\\
Phone: 617-495-8752 \\
Fax: 617-495-7093 \\
email: tgrav@cfa.harvard.edu

\newpage
\section*{Abstract}

We present BVRI colors of 13 Jovian and 8 Saturnian irregular
satellites obtained with the 2.56m Nordic Optical Telescope on La
Palma, the 6.5m Magellan Baade Telescope on La Campanas, and the 6m
MMT on Mt. Hopkins.  The observations were performed between December
2001 to March 2002.  Nearly all of the known irregular satellites can
be divided into two distinct classes based on their colors. One, the
{\bf grey} color class, has the similar colors to the C-type asteroid, 
and the other, the {\bf light red} color class, has colors similar to 
P/D-type asteroids. We also find at least one object, the Jovian 
irregular J~XXIII~Kalyke, that has colors similar to the {\bf red} colored
Centaurs/TNOs, although its classification is unsecure.

We find that there is a correlation between the physical properties
and dynamical properties of the irregular satellites. Most of the 
dynamical clusters have homogeneous colors, which points to single
homogeneous progenitors being cratered or fragmented as the source 
of each individual cluster. The heterogeneous colored clusters are 
most easily explained by assuming that there are several dynamical 
clusters in the area, rather than just one. 

By analyzing simple cratering/fragmentation scenarios, we show that 
the heterogenous colored S~IX~Phoebe cluster, is most likely two
different clusters, a {\bf neutrally} colored cluster centered on 
on S~IX~Phoebe and a {\bf light red} colored cluster centered on
S/2000~S~1. To which of these two clusters the remaining Saturnian
irregulars with inclinations close to $174^\circ$ belong is not
clear from our analysis, but determination of their colors should 
help constrain this. 

We also show through analysis of possible fragmentation and dispersion 
of the six known Uranian irregulars that they most likely make up 
two clusters, one centered on U~XVI~Caliban and another centered on
U~XVII~Sycorax. We further show that, although the two objects have
similar colors, a catastrophic fragmentation event creating one 
cluster containing both U~XVI~Caliban and U~XVII~Sycorax would have 
involved a progenitor with a diameter of $\sim 395$km. While such
an event is not impossible it seems rather improbable, and further
we show that such an event would leave 5-6 fragments with sizes 
comparable or lager than U~XVI~Caliban. Given that the stable region
around Uranus has been extensively searched to limiting magnitudes 
far beyond that of U~XVI~Caliban. The fact that only U~XVI~Caliban
and the larger U~XVII Sycorax have been found leaves us with a 
distribution not compatible with a catastrophic event with such
a large progenitor. The most likely solution is therefore two 
separate events creating two Uranian dynamical clusters.

{\bf Key Words:} Photometry; Satellite of Jupiter; Satellite of Saturn;
Satellite of Uranus; Surfaces, Satellite

\newpage
\section{Introduction}
The satellites of the giant planets in the outer system can be
separated into two groups, the regular and irregular, based on their
orbital characteristics. The orbits of the regular satellites are
prograde, nearly circular, and very close to the equatorial plane of
their host planet. It is commonly accepted that the regular satellites
were created much like the solar system, forming from a
circumplanetary disk of gas and dust. The irregular satellites,
however, have highly inclined and eccentric orbits, with similar
numbers of retrograde as prograde. They are typically small ($\sim
10^2$km) and have semimajor axes that are larger than the regular
satellites. These orbital characteristics suggest that the irregular
satellites formed outside of the circumplanetary disk and were
subsequently captured.

The regular satellites have been studied extensively through both
ground-based and space-based platforms. Aside from the scattered light
from their host planets, the size and brightness of the regular
satellites make them easy targets for even the smallest ground-based
telescopes. The irregulars, however, are sufficiently fainter that,
until recently, the sample of known objects has been severely limited.
N~I~Triton, with its small, circular but retrograde orbit, was
discovered in 1846 by W.~Lassell \citep{Lassell.1846}.  The first distant irregular
satellite, S~IX~Phoebe, orbiting around Saturn, was discovered by
W.~Pickering in 1889.  These, along with eight Jovian (J~VI~Himalia,
J~VII~Elara, J~VIII~Pasiphae, J~IX~Sinope, J~X~Lysithea, J~XI~Carme,
J~XII~Ananke and J~XIII~Leda) and one other Neptunian (N~II~Nereid)
irregular satellites were the only ones known.

Recent discoveries have dramatically increased the census of irregular
satellites.  In 1997-1999, \citep{Gladman.1998,Gladman.2000a} reported
the discovery of five Uranian irregulars. In 1999-2000 a dozen new
irregular satellites were discovered around both Jupiter (J~XVII~Callirrhoe
to J~XXVII~Praxidike and S/2000 J11) \citep{Scotti.2000,Sheppard.2000,Sheppard.2001,Green.2002}
and Saturn (S/2000~S~1-12) \citep{Gladman.2001},
more than tripling the number of known irregulars in the Solar
System. Most recently, the discoveries of another 11 Jovian irregular
satellites (S/2001~J~1-11) \citep{Sheppard.2002} and another
Uranian (S/2001~U~1) \citep{Holman.2002} were reported.

Several scenarios have been proposed for the origin of the irregular
satellites, most concentrating on some type of capture of objects from
heliocentric orbit.  These objects may be temporarily captured in
orbits around the giant planets, by passing through the interior
Lagrange point, $L_2$, with a low velocity \citep{Heppenheimer.1975}.  
The capture is only temporary, since no energy is lost in the transfer
from heliocentric to planetcentric orbit.  Temporary capture typically
lasts only $10-100$ orbits \citep{Byl.1975,Heppenheimer.1977}.  
An energy dissipation mechanism is therefore needed to
make the capture permanent.  The fact that the first seven Jovian
irregular satellites form two dynamical groups, the progrades and the
retrogrades, was recognized early. \citet{Kuiper.1956} first suggested that
these two known classes of Jovian satellites were most likely the
result of two separate events rather than seven (J~XIII~Leda was not
yet discovered). \citet{Colombo.1971} proposed that the capture
was due to fragmentation during a single collision between an outer
Jovian satellite and an asteroid. The fragments of the larger
satellite would comprise the prograde Jovian group, while the
fragments of the asteroid would make up the retrograde group.
\citet{Heppenheimer.1977} studied how a rapid increase of a planet's 
mass during formation could temporarily captured objects into bound
irregular satellites. \citet{Pollack.1979} proposed yet
another capture mechanism. They envisioned that the two clusters were
indeed fragments of separate parent bodies, but that the parent bodies
were captured by gas drag from an extended gas envelope around the
giant planet. They assumed that one pass through the gas envelope was
sufficent to lower the progenitors velocities enough for them to be
permanently captured. The gas envelope collapsed, at least locally,
shortly after the capture of the two progenitors ($\le 10$ years)
ensuring that the gas drag did not cause their orbits to spiral into
the planet. \citeauthor{Pollack.1979} also raised the point that the two
progenitors may have been fragmented after being captured due to the
dynamic pressure of the gas exceeding the progenitor's strength. They
demonstrated that these fragments would stay gravitationally bound to
each other and that a close encounter or collision was needed to
disperse them. Recently \citet{Cuk.2001} have taken the gas drag
model one step further by assuming a flattened disk around a
proto-Jupiter instead of the extended gas envelop used by \citet{Pollack.1979}. 
These theories all imply different time scales and satellite ages.

\citet{Gladman.2001} found that almost all of the newly
discovered irregulars also cluster in easily discernible groups.
They also found more clusters than the two (pro- and retrograde)
initially envisioned. The new Jovian satellites seem to divide the
formerly dispersed retrograde group into at least two subclusters based
on their mean orbital elements.  In addition J~XVIII~Themisto and
J~IX~Sinope (the former prograde, the later retrograde) now appear as
single objects that indicate other yet undiscovered clusters. The Saturnian
irregulars can be divided into three clusters (two prograde and one
retrograde) again based on their mean orbital elements. It is unclear
if S/2000~S~8 falls into the retrograde S~IX~Phoebe cluster. The five
known Uranian irregular satellites have been established to inhabit one
or two retrograde clusters. \citet{Gladman.2001} argue that this
orbital clustering suggests that most of the current irregulars are
fragments from the collisional disruption of earlier captured
progenitor satellites rather than individual satellite captures. A
test of this break-up model was proposed. If the irregulars are
collisional fragments the broad-band colors within a cluster should be
homogeneous.

\vspace{0.5cm}
There exists color photometry, lightcurve, and reflectance spectra for
most of the large irregular satellites of Jupiter and Saturn, but
little photometry has been reported on the recent discoveries. 
\citet{Gladman.2000c} reported BVRI magnitudes of J~XVII~Callirrhoe (formerly
designated S/1999~J~1). BVR
colors of the Uranian irregular satellites U~XVI~Caliban and
U~XVII~Sycorax have been measured by \citet{Maris.2001} and
\citet{Romon.2001}. \citet{Schaefer.2000} observed
N~II~Nereid and determined its UBVRI colors.

Although the physical properties of the larger, previously known
Jovian irregular satellites have been studied in more extensively, the
reported measurements are sometimes contradictory. \citet{Degewij.1980b}
looked at the photometric properties of the outer
planetary satellites and found that the outer Jovian satellites
resemble C-type asteroids. \citet{Degewij.1980a} also measured the
near-infrared reflectance of J~VI~Himalia and S~IX~Phoebe in the $0.3$
to $2.2$ $\mu$m and confirmed that they too resemble C-type
asteroids. \citet{Tholen.1983,Tholen.1984} reported multicolor
observations of six Jovian irregular satellites, as well as S IX
Phoebe, and found that all the progrades resemble C-type asteroids. 
The retrogrades were found to be more diverse in color, with 
J~VIII~Pasiphae and S IX Phoebe looking like C-type asteroids.
J~IX~Sinope was found to be very red, while J~XI~Carme showed a
reflectance spectrum that was flat, but with a strong upturn in the
ultraviolet. \citet{Tholen.1983,Tholen.1984} speculated that J~XI~Carme
might be showing low-level cometary activity with CN emission at
$0.388$ $\mu$m.

\citet{Luu.1991} reported VRI colors, light curves, and reflectance spectra
for J~VI-XIII. She concluded that the satellites resemble a mixture of
C- and D-type asteroids. Similarities between the satellites and the
Trojan asteroids were investigated and found to be consistent with the
hypothesis put forth by \citet{Kuiper.1956} that the two groups of objects
(pro- and retrograde) share a common origin. However, by comparing the
lightcurves of 6 Jovian irregular satellites and 14 Trojan asteroids
\citet{Luu.1991} found that the satellites lack the extreme shapes of the
Trojans. The physical properties of the large Jovian irregular
satellites are generally consistent with, but do not prove, the
capture origin theory proposed by \citet{Pollack.1979}.

In September 1998, six of the then known eight Jovian irregular
satellites were detected in the Two-Micron All Sky Survey \citep{Sykes.2000}. 
The near-infrared colors ($J$, $H$ and $K$) of the
prograde satellites were found to be consistent with the hypothesis
that these objects are fragments of a captured C-type asteroid. The
retrograde satellites in general showed considerable diversity.  The
failure to detect J~XII~Ananke is significant, since it implies a
visible to near-infrared reddening close to that of prograde
J~VI~Himalia and very different from the other retrograde
satellites. This suggests that the retrograde Jovian satellites are
fragments of a more heterogeneous parent body than the prograde
satellites. This result was confirmed by \citet{Rettig.2001}
who performed a BVR color survey of the eight large Jovian irregular
satellites. They noted that the retrograde group appeared redder and
more diverse in both the $B-V$ and $V-R$ colors than the prograde group.

\citet{Gladman.2001} showed that the fragmentation idea proposed 
by \citet{Pollack.1979} was likely incomplete. Capture and subsequent
break-up by gas drag would mean that smaller fragments, which are more
coupled to the gas, should experience faster orbital evolution
spiraling in towards the planet and circularizing their orbits. This
is contradicted by S~IX~Phoebe which is the largest and the inner-most
of its cluster. Also the Uranian U~XVI~Caliban and U~XVII~Sycorax (the
largest irregular satellite) are two of the three closest of their
cluster. \citet{Gladman.2001} favored the idea that each cluster
is produced by disruption of the parent body due to collisional events
well after the capture.

\vspace{0.5cm}
In this paper we present BVRI photometry of a number of the irregular
satellites of Jupiter and Saturn from observations performed at the
Nordic Optical Telescope, MMT Observatory and Magellan Observatory.
We further examine the dispersion of colors and investigate the
correlation between color and dynamical class in order to better
understand the physical and dynamical structure of the irregular
satellites. By comparing the physical and dynamical properties we can
determine wether a dynamical cluster has similar physical properties.
A homogeneous cluster will point to a common progenitor for its members
(the capture of several bodies with similar surface composition is
also a possibility, but this indicate that there are inclinations for
which capture is perfeered, which seems unlikely), 
while a heterogeneous cluster makes additional or other explainations 
necessary. 
 
By determining the orbital and physical properties of the irregular
satellites it is possible to better establish the number of dynamical
clusters around each planet.  Gaining such a census of the number of
clusters, and thus the number of possible progenitors, is necessary to
determine the time scale over which capture of the irregular satellites
happen. This will help distinguish between the different types of
capturing theories.

\section{Observations}

The known irregular satellites have red magnitudes in the range $R
\sim 16-24$, with all the newly discovered objects fainter than $R
\sim 20$.  For the fainter objects medium to large-sized telescopes
are needed to acquire sufficient signal-to-noise ratio to accurately
determine their colors.  We acquired time on three different
telescopes: the 2.51m Nordic Optical Telescope (NOT) on La Palma, the
6.5m MMT on Mt. Hopkins, Arizona and the 6.5m Magellan Observatory at
Las Campanas, Chile.  The circumstances of the observations are given
in Table \ref{tab:obs}.

Most of the observations were performed under photometric conditions
at the NOT using the ALFOSC camera \citep{Sorensen.1996}. The image 
scale of this instrument with its Tektronix 2048x2048
pixel charge-coupled device (CCD) is $0.189$ arcsec per 15$\mu$m
pixel.  A set of broad-band Bessel BVRI filters were used for all the
NOT observations.

A number of observations were also performed under photometric
conditions at the Magellan Baade 6.5m telescope in Chile using MagIC. 
The image scale of this instrument with its SITe
2048x2048 pixel, 4 amplifier CCD is $0.069$ arcsec per 24$\mu$m
pixel. The MagIC has a set of Harris BVRI filters that were used for
the observations.

Additional observations were performed under photometric conditions at
the MMT 6.5m telescope on Mt.  Hopkins, Arizona, using the MiniCam
\citep[a protoype of Megacam;][]{McLeod.2000}. The Minicam has two 2048x4608 pixel
CCDs, which in 2x2 binned mode, gives $0.091$ arcsec per $27\mu$m
pixel. The two chips are mounted side by side with a 1mm ($3.5$
arcsec) gap. The MMT has a set of Harris BVRI filters that were used
for the observations.

The irregular satellites move across the sky at rates of $5-12$ arcsec
per hour.  Thus, the exposure time were limited to avoid trailing
loss. Since photometry of field stars is needed to do aperture
photometry (see next section), we need to track at sidereal rates,
rather than tracking at the satellites' rates. With seeing of
$1.0-1.5$ arcseconds, the maximum exposure time is limited to $240$s
for a Jovian irregular satellite and to about $300$s for a
Saturnian. Several images were therefore taken in each filter. The images
were then shifted and co-added to increase the signal-to-noise ratio and 
to avoid trailing losses.

\section{Data Reduction}

All the images were overscan corrected, bias subtracted, and flat
field corrected to remove differences in the sensitivity over the CCD. 
The bias was created by averaging 10-15 zero-time exposure. The flatfield
frames were created by taking the median of a series of short
exposures of the evening/morning twilight or of an illuminated white
screen inside the telescope dome. This removed most of the
irregularities found on the CCD, producing very uniform images with
less than $1$\% variability across the CCD. 

The photometry was flux calibrated using numerous standard stars
\citep{Landolt.1992}. Extinction coefficients and transformation equations
were determined for each night. The DAOPHOT package \citep{Stetson.1990} 
in IRAF \citep{Tody.1986,Tody.1993} was used to determine the
instrumental magnitudes, transformation equations, photometry, and
colors of the objects. 

Since many of the satellites are very faint and have errors dominated
by sky noise, the technique of aperture correction was used 
\citep{Howell.1989}. The aperture sizes were picked to ensure that 
the inner aperture covered about one and a half times the full-width at half maximum
(FWHM) of the psf of the object. This aperture size insure that a maximum
signal-to-noise ratio is achieved for the objects \citep{Dacosta.1992}.
Thus, a aperture radius of 9 pixels was used for the data taken at NOT and 
MMT ($\sim1.7$ and $\sim 0.9$ arcseconds respectivelly), while 12 pixels ($\sim
0.9$ arcseconds) was used for the Magellan observations. For the outer
aperture 2-3 times the FWHM of the psf was used and experimentation
showed that 25 pixels worked well for all the data. For the
determination of the sky background an annulus 10 pixels wide with an
inner diameter of 30 pixels was used. The aperture correction
magnitude varies with seeing, but is generally in the range $\Delta m
\sim 0.1-0.3$ with errors of$\pm 0.01-0.03$.

In some of the images taken using the I-band filter, severe fringing
effects made reduction of the images difficult. Since the fringing
effects were confined to the outer parts of the CCD, only the images
where the positional uncertainties of the target were great did it
effect the data reduction. By summing the available images taken
using the I-band filter it was possible to create a fringing flat that
removed most of the fringing, resulting in images with less than $3$\%
variability across the CCD. 

\section{Photometric Results}

In this survey BVRI photometry and the associated colors were
determined for 13 Jovian and 8 Saturnian irregular satellites, out of
the 20 Jovian and 13 Saturian irregular satellites known at the time
of observation. Since most of the observations were performed at the
2.51m Nordic Optic Telescope at La Palma, the survey is mostly limited
to the objects brighter than $m_V = 23.0$.

Table \ref{tab:res} gives the absolute magnitude, normalized to unit
heliocentric and geocentric distance and zero degree phase angle
\citep{Bowell.1989}, the observed $V$-magnitude, and the $B-V$, 
$V-R$, and $V-I$ colors with their associated errors. The corresponding
solar colors are approximately $B-V = 0.67$, $V-R = 0.37$ and $V-I =
0.71$ \citep{Hartmann.1982}.

The $B-V$ vs $V-R$ colors are shown in Fig. \ref{fig:bvvr} similar
to the plots that have become the norm for identifying structures in
the color distribution. \citet{Rettig.2001} used a similar
plot to show the difference in colors between the prograde and
retrograde populations of the Jovian irregular satellites. Figures
\ref{fig:bvvr}-\ref{fig:vrvi} includes all the objects observed in
this survey, as well as colors reported by for J~VI~Himalia,
J~VII~Elara, J~XIII~Leda \citep{Rettig.2001} and the two
Uranian U~XVI~Caliban and U~XVII~Sycorax \citep{Maris.2001,Romon.2001}. 
Figure \ref{fig:bvvrtno} also includes the
Centaurs and TNOs for which $B-V$ and $V-R$ colors with uncertainties
below $0.05$ magnitudes are available.

Figure \ref{fig:bvvr} confirms the distribution of colors found by
\citet{Rettig.2001}.  The objects can be roughly separated into
two distinct classes based on their colors. One which is essentially
{\bf grey} in color (the objects have sun-like colors) and one which
will be hereafter called {\bf light red}. The $B-V$ vs. $V-I$ plot in
Fig. \ref{fig:bvvi} and $V-R$ vs. $V-I$ plot in Fig.
\ref{fig:vrvi} confirm this distribution. Note that one object changes
classification between the plots. We attribute this to an
underestimation of the error of the I-band photometry due to fringing
effects. It should also be noted that there exists few estimates of
variation in the colors due to variation in surface properties.
S~IX~Phoebe is a good example of an irregular satellite that has a
large bright spot on one side of it's surface \citep{Simonelli.1999}. 
How this bright spot affects the color as S IX Phoebe rotates
has not yet been investigated.

These two classes should not be confused with the two different
classes named {\bf grey} and {\bf red}, found in the Centaur/TNO
population by \citet{Tegler.1998,Tegler.2000}. Figure
\ref{fig:bvvrtno} shows that both the {\bf grey} and {\bf light red}
classes as defined in this paper coincide with what \citet{Tegler.1998,Tegler.2000}
call the {\bf grey} class.

\section{Dynamical Class and Colors}

When comparing colors of objects within a dynamical cluster it becomes
evident that for most clusters the individuals members have similar
colors, suggesting a common progenitor.  Figure \ref{fig:polar} shows
a polar plot of the mean semimajor axis $a$, normalized to the Hill
sphere, $R_H$, for each parent planet, and mean inclination $i$
\citep[;JPL Solar System Dynamics website ({\it ssd.jpl.nasa.gov})]{Pollack.1979,Gladman.2001}. 
The different Jovian dynamical clusters
are immediately apparent, while the more disperse Saturnian and
Uranian clusters are harder to spot. Objects for which photometry has
been performed have been colored according to their apparent color
class, {\it blue} for {\bf grey} (C-type) colored objects and {\it
red} for {\bf light red} (P- or D-type) colored objects. The
unobserved objects have black symbols.
  
Assuming that the most massive body in each cluster is close to the
original orbit before a catastrophic collision, the colored areas in
Fig. \ref{fig:polar} show the possible dispersion of fragments in
semimajor axis and inclination.  We generated these dispersion areas
by adding fixed length velocity vectors with random directions to the
velocity vector of the largest member of each cluster, using that
objects mean orbital elements and assuming that the collision occurs
at a given orbital phase and is an relative small impulse.  Given
for each resulting velocity vector, new orbital elements are
calculated.  The cluster of points can be understood in the limit of
small velocity changes.  We assume typical dispersion velocity $\Delta
{\bf v} = \Delta {\bf v}_S + \Delta {\bf v}_T + \Delta {\bf v}_W$,
where the velocity vector has been split into three components.
$\Delta {\bf v}_W$ gives the component perpendicular to the orbital
plane of the satellite, while $\Delta {\bf v}_S$ and $\Delta {\bf
v}_T$ are the components along and at right angles to the radius
vector, ${\bf r}$ of the satellite from the planet. The dispersion in
$a$-$i$ phase space is then given by \citep{Roy.1988}
\begin{equation}
   \Delta a = \frac{2}{n \sqrt{1-e^2}} \left( e \sin f \Delta v_s +
   \frac{p}{r} \Delta v_T \right)
\end{equation}
and
\begin{equation}
   \Delta i = \frac{r \cos u}{na^2 \sqrt{1-e^2}} \Delta v_W
\end{equation}
where $f$ is the true anomaly, $p = a(1-e^2)$ and $u = f +
\bar{\omega} - \Omega = f - \omega$.

All of the known clusters are made up of one large and several smaller
bodies. The smallest mass difference between the largest and next largest
bodies can be found in the Saturnian $45^\circ$-inclination cluster
(mass-ratio $\sim 5$) and the potential Uranian cluster containing both
U~XVII~Sycorax and U~XVI~Caliban (mass-ratio $\sim 7$). Evaluating the 
center of mass for these two show that dispersion areas centered on the
{\it ``center of mass''} are near identical to the dispersion areas centered
on the largest body of the cluster. 

The speed of the fragments and the true/mean anomaly of the progenitor
at impact have been adjusted to show plausible divisions of the
objects into clusters.  While a more careful analysis that accounts
for post-collisional orbital evolution and the probability of impact
at different orbital phase is beyond the scope of this paper, the
scatter plots provide an zeroth-order approximation of the dispersion velocities
needed to create the clustering structure seen in the irregular
satellites around all the giant planets. 

\vspace{0.5cm}

\citet{Rettig.2001} determined colors for the 8 brightest
Jovian irregular satellites and noted that the colors for the
retrograde were less homogeneous and generally redder than those of
the four prograde irregular satellites. In this survey we have
observed 7 more retrograde and one prograde, as well as J~VIII~Pasiphae, 
J~IX~Sinope, J~XI~Carme and J~XII~Ananke. We find that
the retrograde dynamical cluster centered on $149^{\circ}$ inclination
is indeed severely heterogeneous, while the cluster around J~XI~Carme
is homogeneous with objects in the {\bf light red} color class.

We also observed eight Saturnian irregular satellites and found that
the two prograde clusters centered on $34^{\circ}$ and $46^{\circ}$
were found to be homogeneous with {\bf light red} colors like the
D-type asteroids. The retrograde cluster containing S~IX~Phoebe was
however heterogeneous, with one {\bf grey} colored object
(S~IX~Phoebe) and one with {\bf light red} color (S/2000 S 1).

\vspace{0.5cm}

In the following sections we examine in more detail the different
clusters and propose some possible theories for the origin of each
dynamical cluster.

\subsection{The Jovian Prograde ($28^\circ$ inclination) cluster}

The Jovian prograde $28^\circ$ inclination-cluster are all clustered in
the {\bf grey} color class with weighted mean colors $B-V = 0.66 \pm
0.02$ and $V-R = 0.36 \pm 0.01$. Of the five known members of this
cluster only J~X~Lysithea was observed during this survey, while the
colors of J~VI~Himalia, J~VII~Elara and J~XIII~Leda were taken from
\citet{Rettig.2001}. We did not observe S/2000~J~11, which
remains the only Jovian irregular satellite discovered in 2000 that
has not been recovered. The colors of J~X~Lysithea reported in this
paper are consistent with colors reported by \citet{Luu.1991} and 
\citet{Rettig.2001}. The {\bf grey} colors of the prograde cluster imply
that they have surfaces similar to that of C-type asteroids. 
\citet{Cuk.2001}, studying the gas-assisted capture of irregular
satellites, have found that the progenitor of the prograde cluster may
have originated from the Hilda region, about $4$AU from the Sun. The
Hilda family is dominated by P- and D-type asteroids, but due to a
spectral slope - asteroid size correlation, the largest members of the
family 153 Hilda (with a diameter of $170$km) and 334 Chicago (with a
diameter of $156$km) are possibly C-type asteroids \citep{Dahlgren.1997}.  
It is usually assumed that the progenitor of a
fragmented dynamical cluster had roughly twice the size of the clusters
largest member today. J~VI~Himalia with its diameter of $\sim85$km
implies a progenitor body of similar size as 153 Hilda and 334
Chicago. \citet{Jarvis.2000a} however classify J~IV~Himalia as
an F-class asteroid, a subclass of C, and propose that it originated
as part of the Nysa asteroid family. The spectroscopic data collected
by \citet{Jarvis.2000a} however show evidence of absorption
features at 0.7 $\mu$m, which is only weakly or non-evident in other
F-class asteroids \citep{Sawyer.1991}.

\vspace{0.5cm}

Figure \ref{fig:polar} shows the a-i phase space dispersion with a
fragmentation speed of $150$ m/s centered on J~VI~Himalia, assuming
that the collisional event happened at pericenter. This dispersion
speed is approximately twice the escape velocity, at $v_e \sim 102$m/s,
and clearly indicates that the five objects in the vicinity are
fragments of a common progenitor, that most likely originated in the
outer part of the Main Asteroid Belt.

\subsection{The Jovian IX~Pasiphae and XII~Ananke ($149^\circ$ inclination) cluster(s)}

The $149^\circ$ cluster is currently the group with the greatest number
of members. This cluster is dynamically the most diverse of the known
Jovian clusters with inclinations ranging from $147.1^\circ$ to
$152.7^\circ$. This is a factor of 3 larger than the dispersion of the
prograde $28^\circ$ cluster and a factor of ten larger than the
dispersion of the $165^\circ$ cluster. The $149^\circ$ cluster also has
the largest range of semimajor axis, about 3 times more dispersed than
the $28^\circ$ cluster and 6 times more than the $165^\circ$ cluster. The
orbital mean elements suggest two distinct dynamical clusters instead of
one, one centered on J~IX~Pasiphae and one centered on J~XII~Ananke.
As can be seen in Fig. \ref{fig:polar} there appears to be a gap in 
the semimajor axis between the two clusters. J~VIII~Sinope, with its mean 
orbital inclination of $158.1^\circ$, seems to be an independent capture 
of an object that has not been fragmented.

This paper presents colors for all seven members of this cluster that
were known at the time of observation, but for J~XXII~Harpalyke the $B-V$
color could not be determined, due to its faintness. The results of
the observations can be seen in Table \ref{tab:res} and Fig.
\ref{fig:bvvr}-\ref{fig:vrvi}. The cluster has a very disperse set of
colors with weighted mean $B-V = 0.75\pm0.04$ and $V-R =
0.41\pm0.08$. Both J~IX~Pasiphae and J~XII~Ananke have been observed
earlier \citep{Degewij.1980c,Tholen.1984,Luu.1991,Rettig.2001}. 
The observations of J~IX~Pasiphae given in
this paper are in good agreement with these previous observations and
the weighted mean colors of all available observations give the colors
$B-V = 0.71 \pm 0.05$, $V-R = 0.39 \pm 0.02$ and $R-I = 0.75 \pm
0.03$. This places it in the {\bf grey} color class.

Our observations of J~XII~Ananke are consistent with other
observations in the $V-R$ color, but our $B-V$ color is slightly high
at $B-V = 0.90 \pm 0.06$ compared to $0.76 \pm 0.03$ (Rettig {\it et
al} 2001). Our results differ from those of Rettig {\it et al} only at
the $2\sigma$ level and could be the result of slight contamination by
background source or statistical fluctuations. The weighted mean
colors of all the observations give $B-V = 0.77 \pm 0.03$, $V-R = 0.42
\pm 0.02$ and $V-I = 0.83 \pm 0.04$, and places J~XII~Ananke right on
the border between the {\bf grey} and {\bf light red} color
classes. Further observations with higher accuracy are needed to
determine the color of this object, but currently the observations
seems to favor it as a {\bf light red} object, rather than {\bf grey}.

\vspace{0.5cm}

We are thus left with a rather confusing collection of orbits and
colors.  J~IX~Pasiphae has a {\bf grey} color, while the irregular
satellites with similar orbits (J~XVII~Callirrhoe and J~XIX~Magaclite) have {\bf
light red} colors. For the objects clustered around J XII Ananke the
problem is the same, but reversed, with J~XII~Ananke exhibiting a {\bf
light red} color and J~XXII~Harpalyke, J~XXVII~Praxidike and J~XXIV~Iocaste having {\bf
grey} colors. It should again be stressed that the color of
J~XII~Ananke is right on the border between the two classes of
colors. Further and more accurate observation is necessary to
determine its color beyond doubt. \citet{Saha.1993} demonstrated
that both J~VIII~Sinope and J~IX~Pasiphae are trapped in secular
resonances. It is reasonable to assume that significant orbital
evolution has moved the object in this area of phase space away from
their imidiate post-fracture orbits, trapping some in the secular
resonances. A more detailed study of both the physical and dynamical
properties of these bodies are needed to better understand the
mechanism that govern the evolution of these bodies.

It should be noted that it is possible to find dispersion ellipses,
using dispersion velocities of $200$m/s, that create homogeneous
clusters. This however create rather arbitrary combination of objects
that does not seem to have anything but colors in common. 

\subsection{The Jovian XI Carme ($165^\circ$ inclination) cluster}

The objects of the Jovian XI~Carme cluster has very little dispersion in
their mean orbital elements of semimajor axis and inclination. Three
of the six known objects from this cluster were observed during this
survey. Two, J~XI~Carme and J~XX~Taygete, lay in the {\bf light red} color
class, while J~XXIII~Kalyke has colors more consistent with the {\bf red}
class defined by \citet{Tegler.1998,Tegler.2000}.  J~XXIII~Kalyke is the
only object in the sample to exhibit this redness. But looking at the
$V-I$ color of J~XXIII~Kalyke it is seem to be part of the {\bf light red}
class. Why J~XXIII~Kalyke has such a red $V-R$ color, while it's $V-I$
color is moderate is uncertain, but it is possible that the moderate
$V-I$ color is due to underestimated fringing effects. The observation
of J~XXIII~Kalyke at the NOT on Jan 24. 2002, indicate that the $V-I$
color could be high ($0.96 \pm 0.16$). Two separate observation was
taken of J~XXIII~Kalyke to confirm the high $V-R$ color and both are
consistent within the estimated errors, but further observations are
needed to confirm the {\bf red} color.

\vspace{0.5cm}

The J~XI~Carme cluster is consistent with a collisional event giving the
fragments speeds at $\sim 30$m/s, which is just beyond the
escape velocity of J~XI~Carme ($v_e \sim 28$m/s). It is by far the
thightest of the dynamical clusters and seems to have been very little
post-fracture orbital evolution. We propose that the J~XI~Carme cluster
has a D-type progenitor, most likely from the Hilda or Trojan
families. How J~XXIII~Kalyke fit's into the cluster is still very much an open
question. The {\bf red} color points to an origin in the Centaur/TNO
population and it is possible that J~XXIII~Kalyke is a surviving fragment
of the impactor that caused the fragmentation of the J~XI~Carme
progenitor, allthough it seems very unlikely that such a fragment would
have a post-collisonal orbit exactly that of the progentior it impacted.

\subsection{The Saturnian $34^\circ$ and $46^\circ$ inclination clusters}

The two prograde Saturnian clusters, with mean orbital inclinations
around $34^\circ$ and $46^\circ$ also have fairly tightly clustered
colors. They both lay in the {\bf light red} class. The
$34^\circ$-cluster has weighted mean colors of $B-V = 0.91 \pm 0.05$
and $V-R = 0.48 \pm 0.01$. The moderate dispersion in the $B-V$ color
is due to the high uncertainty in the measurements of this color in
all the three objects in this cluster that was measured. The
$46^\circ$-cluster has weighted mean colors of $B-V = 0.79 \pm 0.01$
and $V-R = 0.51 \pm 0.02$.

It is important to mention that although the two clusters seem to be
homogeneous, there are significant uncertainties. S/2000~S~10,
belonging to the $34^\circ$ inclination cluster, is clearly in the
{\bf light red} color class when looking at the $B-V$ vs. $V-R$ plot
in figure \ref{fig:bvvr}. It has however a very low $V-I$ color at
$0.61 \pm 0.12$, which puts it in the {\bf grey} color class in the
$B-V$ vs. $V-I$ (figure \ref{fig:bvvi}) and $V-R$ vs. $V-I$ (figure
\ref{fig:vrvi}) plots.  This is most likely due to the combination of
the faintness of S/2000~S~10, ith a V magnitude at $23.90 \pm 0.05$,
and the fringing effects that show up in the $I$ filtered images when
doing long exposures. It is likely that the formal errors for the
$V-I$ color of S/2000~S~10, as given in table \ref{tab:res}, is
underestimated. We have therefore decided to disregard the $V-I$ color
of this object and classify it as a {\bf light red} colored
object. More observations are necessary to confirm this classification.

\vspace{0.5cm}

Both these dynamical clusters are neatly confined. All the objects in
the $34^\circ$ inclination cluster easily fit inside the dispersion
ellipse given fragment speeds of $50$m/s centered on S/2000~S~11,
which is the biggest member of the cluster with a diameter of
13~km. Using this diameter, S/2000~S~11 has an escape velocity of only
$\sim 5$m/s.

While the $46^\circ$ inclination cluster is physically homogeneous, a
large dispersion velocity is needed, at $350$m/s centered on its
largest member, S/2000~S~3, to be able to fit a dispersion area that
contains them all. S/2000~S~3, with a diameter of 16~km, has an escape
velocity of $v_e \sim 18$m/s. The large dispersion velocity could be
due to a close encounter with a large object that was subsequently
cleared away. It was recently reported that there has been discovered
possible secular resonant motion in three of the four known members of
this dynamical cluster \citep{Carruba.2002a,Cuk.2002}. These secular 
resonances may have caused significant
post-fragmentation orbital evolution that could be behind the large
dispersion of orbital elements in this cluster.

\subsection{The Saturnian IX Phoebe ($174^\circ$ inclination) cluster}

Only two of the four known members of the $174^\circ$ inclination
cluster were observed. S~IX~Phoebe was found to lay in the {\bf grey}
class, while S/2000~S~1 is part of the {\bf light red} color class.

The color of S~IX~Phoebe found here is consistent with $B-V$
observations from \citet{Degewij.1980c} and \citet{Tholen.1983}. 
\citet{Degewij.1980c} also reported a single $V-R$
observation of $0.66\pm0.02$, which differs significantly from the
color found in this survey. Spectral observations of S~IX~Phoebe in
the $0.33-0.92$ $\mu$m region at two subsequent nights by 
\citet{Buratti.2002} both show essentially flat and grey spectra, which is
consistent with the color observations in this paper.

\vspace{0.5cm}
The size distribution of the S~XI~Phoebe cluster, with one large and
several small bodies, points to one or more cratering events. There is
an upper limit on the energy density of a collision for a cratering
event, given by $E/M_T \sim 10^2$J/kg, where $E$ is the kinetic energy
of the impactor and $M_T$ is the mass of the target \citep{Fujiwara.1989}.
Energies larger than this will lead to fragmentation of the target,
leaving a different mass distribution than found in the S~IX~Phoebe
cluster.  Assuming that such event only knocked loose a piece of similar
size of S/2000~S~1 ($\sim 7 \cdot 10^{-4}M_T$) and that $10\%$ of the
energy is converted into kinetic energy for the piece \citep{Fujiwara.1989}, a
maximum speed of $\sim 170$m/s is achieved. For objects with similar
sizes as S/2000~S~12 ($\sim 3 \cdot 10^{-6}M_T$), using the same
arguments, yield a maximum dispersion speed of $\sim 2.6$km/s.  It is
important to remember that these are generous upper limits, assuming
that the maximum energy possible for a cratering event and also
assuming that only one piece is knocked out, with all the kinetic
available being used in accelerating it. If we assume that both
S/2000~S~7 and S/2000~S/12 (for a total mass of $\sim 6 \cdot
10^{-6}M_T$) were part of the same event, we get upper dispersion
velocities of $\sim 1.8$km/s.

Maximum dispersion in semimajor axis is achieved when the
fragmentation/cratering happens at or close to the orbits pericenter.
Dispersion speeds of over $300$m/s is needed to include S/2000~S~1 in
a dispersion area centered on S~IX~Phoebe.  It is very unlikely that
it is a piece that has been carved of S~IX~Phoebe due to cratering.
The {\bf light red} surface color of S/2000~S~1 further supports this,
meaning that S/2000~S~1 probably is the largest remaining piece of a
catastrophic fragmentation of another progenitor. On the other hand,
it is possible to create a dispersion ellipse using a dispersion
velocity of $\sim 250$m/s, that includes both S/2000~S~7 and
S/2000~S~12. By increasing this to dispersion velocity to $\sim
350$m/s and assuming that the cratering happened at apocenter rather
than pericenter S/2000~S~9 can be included as well.  It is therefore
possible that these three irregular satellites are pieces from a
cratering event on S~IX~Phoebe.  While this dispersion velocity is
high enough to create dispersion areas that include S/2000~S~1, it is
not included in this case as the cratering happens at apocenter it
results in a larger dispersion in inclination and thus reducing the
dispersion in semimajor axis.  It is also possible, using dispersions
velocities of $\sim 250$m/s and an event close to $M=220^\circ$, to
create a dispersion area centered on S/2000~S~1 that includes all
three of the irregular satellites, S/2000~S~7, S/2000~S~8 and
S/2000~S~12.

We are therefore left with the conclusion that there is are two rather
than one dynamical cluster, both with inclinations $\sim
174^\circ$. It is however not clear to which of these two clusters the
three small irregular satellites, with diameters of$\sim 2-4$km, that
reside in orbits between S~IX~Phoebe and S/2000~S~1 belong
to. Determining their colors would help determine which of the two
progenitors they most likely are the result of.

It should also be noted that it is possible that orbital evolution
after fragmentation of {\bf one single} progenitor is behind the
current orbital distribution. This suggest that S~IX~Phoebe either
moved inwards or S/2000~S~1 diffused outwards.  Both these scenarios
seems unlikely as no secular resonances have been reported to be
present in this region around Saturn. The fact that the two objects
observed in this survey have such different colors, also support the
idea of two rather than one dynamical cluster in the vicinity.

\subsection{The Uranian Irregular Satellites}

The six known Uranian irregular satellite all have similar orbital
elements, but with a rather large dispersion. It has been
speculated that these irregulars make up several dynamical clusters,
but no clear evidence of this has been found. Only U~XI~Caliban and
U~XII~Sycorax have had their $B-V$ and $V-R$ colors determined 
\citep{Maris.2001,Romon.2001}, and both have {\bf light red} colors.

Looking at the possible dispersion caused by fragmentation makes it
clear that there is likely more than one dynamical cluster, unless the
irregulars have had significant orbital evolution after
fragmentation. Using dispersion velocities of $75$ and $115$m/s
centered on U~XVI~Caliban ($v_e \sim 45$m/s) and U~XVII~Sycorax ($v_e
\sim 87$m/s), respectively, the six Uranian are divided into two
clusters (see Fig. \ref{fig:polar}).  U~XVIII~Prospero and
U~XIX~Setebos are then part of the U~XVII~Sycorax cluster, while
U~XX~Stephano falls in the U~XVI~Caliban cluster. The new S/2001~U~1
does not fit into either cluster, but it's current mean orbital elements
(which are still rather uncertain)
suggest that it is part of the U~XVII~Sycorax cluster.  It is possible
to fit all the Uranian irregulars inside a dispersion area centered
on U~XVII~Sycorax (the biggest of the Uranian irregulars with a
diameter of $\sim 95$km), using a dispersion velocity of $\sim
250-300$m/s and a collisional event close to apocenter. To create such
a dispersion velocity for U~XVI~Caliban alone, a fragmentation caused
by a event with energy density $E/M_T \sim 5 \cdot 10^4$J/kg is needed
(again assuming that $10\%$ of the energy of the impact is used to
disperse the fragments). 

Using this energy density it is possible to estimate the mass of the
target \citep{Davis.1989}
\begin{equation}
   f_l = 0.5 \left( \frac{2SM}{\rho E} \right)^{1.24}
\end{equation}
where $S$, $M$ and $\rho$ is the impact strength, mass and density of
the target respectively, $E = (1/2) m v^2$ is the kinetic energy of the
impactor and $f_l$ is the mass ratio of the largest remaining fragment
to the target. Using U~XVII~Sycorax as the largest remaining fragment
and the energy needed to disperse U~XVI~Caliban to its present
position we get a progenitor with a diameter of $\sim 395$km, assuming 
the progenitor was an icy body with impact strength $S/\rho \sim 30 J/kg$.
The mass ratio of U~XVII~Sycorax to the progenitor is $f_l \sim 0.16$. 

This implies that the Uranian irregular satellites could be the
remains a fractured natural satellite or a large captured
Centaur/Trans-Neptunian Object. However, assuming a size distribution
given by power-law \citep{Greenberg.1978}
\begin{equation}
   N(>m) = C m^{-b}
\end{equation}
where $N$ is the number of fragments with mass larger than $m$, and
$C$ and $b$ are constants that can be determined through the total
mass of the target $M$ and the size of the largest fragment. Applying
this to the Uranian system yields a distribution with 5-6 
fragments with mass equal to or larger than U~XVI~Caliban. The stable
region around Uranus has been searched to a $\sim 90\%$ efficency
to a $50$\% ``completeness magnitude'' of $m_R \sim 24.3 - 25.1$ 
\citep{Gladman.1998,Gladman.2000a}. Given U~XVI~Calibans magnitude, 
$m_R = 21.9 \pm 0.1$ \citep{Gladman.1998}, it is hard to
imagine that the ``missing'' three or four fragments would have been missed in 
these searches. The one dynamical cluster hypothesis therefore looks
unlikely.

\section{Conclusion}

The observations presented in this paper provide a survey of BVRI
magnitudes and colors for 13 Jovian and 8 Saturnian irregular
satellites. The results show that the irregular satellites seem to be
divided into two classes based on colors. To avoid adding confusion
with other terminology used in the field, the names of these two
classes of have been chosen as {\bf grey} and {\bf light red}.
Observations of additional objects are needed to confirm this
division.

The colors of the irregular satellites found in this survey confirm
that both C- and P/D-type asteroids are the most likely captured
parent bodies, although at leat one (J~XXIII~Kalyke) has colors too red
for it to be a P/D-type asteroid. This indicates that it is possible
that one or more Centaur/TNO have also been involved in the capture
and fragmentation process, or some of the irregular satellites have
experienced significant surface evolution.

The data also show a correlation between physical and dynamical
properties in most of the clusters. This suggest that most clusters
are the products of cratering or catastrophic fragmentation events. We
also found that two of the dynamical clusters (the Jovian $145^\circ$
and the Saturnian Phoebe cluster) have heterogeneous colors. The
diverse and highly populated Jovian $145^\circ$ cluster seems to most
likely be the result of two (or maybe even three)
cratering/fragmentation events.  The existence of secular resonances
in the region \citep{Saha.1993,Carruba.2002a,Cuk.2002} is good
evidence that significant orbital evolution has occured after the
cratering/fragmentation events.

The S~IX~Phoebe cluster was also found to be heterogeneous and we show
through a simplified analysis of possible cratering/fragmentation
scenarios and subsequent dispersion velocities that these fragmets are
most likely two rather than one dynamical cluster, one centered on
S~IX~Phoebe and another centered on S/2000~S~1. It is from this
analysis not possible to determine which of these two clusters the
other Saturnian irregular satellites in the area (S/2000~S~7,
S/2000~S~9 and S/2000~S~12) originated from. Study of the physical
properties of these three satellites should help determine if they are
part of the {\bf light red} colored S/2000~S~1 cluster or the {\bf
neutral} colored S~IX~Phoebe cluster. It should also be noted that
\citet{Buratti.2002} suggest that the dark leading side of
Iapetus and the surface of Hyperion are being coated by particles from
retrograde satellites exterior to Iapetus' orbit.  They determined
that the visual spectra taken of the two surfaces can be explained as
a linear admixture of two components: an icy satellite in the
Saturnian system and a D-type asteroid.  S~IX~Phoebe's visual spectral
properties are unlike those of either Iapetus or Hyperion, so
S~IX~Phoebe does not apparently contribute to the darking of Iapetus's
leading side.  However, the colors determined for S/2000~S~1 show that
it is consistent with a D-type asteroid and could therefore play a
role in the darkening of Iapetus.

We also used the simple cratering/fragmentation analysis and showed
that the six known Uranian irregular satellites can be divided into
two dynamical cluster, one centered on U~XVI~Caliban and another on
U~XVII~Sycorax, although there is an unlikely possibility that all
six are fragments of a catastrophic fragmentation of a large (diameter
of $\sim 430$km) natural moon or captured Centaur/TNO.

\vspace{0.5cm}

In future studies we will determine BVRI colors of the irregular
satellites not observed in this paper in order to help distinguish 
between the different dynamical clusters. Determining the colors of the 
remaining known Saturnian irregulars with inclinations around $174^\circ$ 
is especially important. We will also investigate the
near-IR JHK-bands to extend the spectral baseline, which will help in 
distingushing the surface properties between objects with similar BVRI
colors. This will help to classify the large and confusing set 
of Jovian irregulars in the $140^\circ - 160^\circ$ inclination area.

\section{Acknowledgments}

This work has been supported by a Smithsonian Astrophysical
Observatory Pre-doctoral Fellowship at the
Harvard-Smithsonian Center for Astrophysics, Cambridge, USA.

We thank Brian Marsden and Gareth Williams of the Minor Planet Center
for providing ephemerides, and Kris Stanek for his
willingness to answer questions about photometry.

T. Grav was a visiting astronomer at the Nordic Optical Telescope. The
Nordic Optical Telescope is operated on the island of La Palma jointly
by Denmark, Finland, Iceland, Norway and Sweden, in the Spanish
Observatorio del Roque de los Muchachos of the Instituto de
Astrofisica de Canarias.  Part of the data presented here have been
taken using ALFOSC, which is owned by the Instituto de Astrofisica de
Andalucia (IAA) and operated at the Nordic Optical Telescope under
agreement between IAA and the NBIfAFG of the Astronomical Observatory
of Copenhagen.

T. Grav and M.Holman were visiting astronomers at the MMT. The MMT is 
a joint facility owned by the Smithsonian Institution and the University 
of Arizona, and is located on Mt. Hopkins, Arizona.  

T. Grav and M.Holman were visiting astronomers at the Magellan Baade
Telescope. The Magellan Observatory is a collaboration between the 
Carnegie Observatories, the University of Arizona, Harvard University, 
University of Michigan, and MIT. The Magellan Observatory are located 
at Las Campanas Observatory, Chile, and operated by OCIW.

---------------------------------------------------------------
\newpage

\bibliography{satcolor}

\begin{thebibliography}{}

\bibitem[{Bowell} {\em et~al.}(1989){Bowell}, {Hapke}, {Domingue}, {Lumme},
  {Peltoniemi}, and {Harris}]{Bowell.1989}
{Bowell}, E., B.~{Hapke}, D.~{Domingue}, K.~{Lumme}, J.~{Peltoniemi},\ and
  A.~W. {Harris} 1989.
\newblock {Application of photometric models to asteroids}.
\newblock In {\em Asteroids II}, pp.\  524--556.

\bibitem[{Buratti} {\em et~al.}(2002){Buratti}, {Hicks}, {Tryka}, {Sittig}, and
  {Newburn}]{Buratti.2002}
{Buratti}, B.~J., M.~D. {Hicks}, K.~A. {Tryka}, M.~S. {Sittig},\ and R.~L.
  {Newburn} 2002.
\newblock {High-Resolution 0.33-0.92 {$\mu$}m Spectra of Iapetus, Hyperion,
  Phoebe, Rhea, Dione, and D-Type Asteroids: How Are They Related?}
\newblock {\em Icarus\/}~{\bf 155}, 375--381.

\bibitem[{Byl} and {Ovenden}(1975){Byl} and {Ovenden}]{Byl.1975}
{Byl}, J.,\ and M.~W. {Ovenden} 1975.
\newblock {On the satellite capture problem}.
\newblock {\em \mnras\/}~{\bf 173}, 579--584.

\bibitem[{Carruba} {\em et~al.}(2002){Carruba}, {Burns}, {Nicholson}, {Cuk},
  and {Jacobson}]{Carruba.2002a}
{Carruba}, V., J.~A. {Burns}, P.~D. {Nicholson}, M.~{Cuk},\ and R.~A.
  {Jacobson} 2002.
\newblock {S2000S5 and S/2000S6: Saturnian moons trapped in the Kozai
  resonance}.
\newblock {\em AAS/Division for Planetary Sciences Meeting\/}~{\bf 34}.

\bibitem[{Colombo} and {Franklin}(1971){Colombo} and {Franklin}]{Colombo.1971}
{Colombo}, G.,\ and F.~A. {Franklin} 1971.
\newblock {On the formation of the outer satellite groups of Jupiter}.
\newblock {\em Icarus\/}~{\bf 15}, 186--189.

\bibitem[{C\'uk} and {Burns}(2001){C\'uk} and {Burns}]{Cuk.2001}
{C\'uk}, M.,\ and J.~A. {Burns} 2001.
\newblock {Gas-Assisted Capture of the Irregular Satellites of Jupiter}.
\newblock {\em AAS/Division for Planetary Sciences Meeting\/}~{\bf 33}.

\bibitem[{C\'uk} {\em et~al.}(2002){C\'uk}, {Burns}, {Carruba}, {Nicholson},
  and {Jacobson}]{Cuk.2002}
{C\'uk}, M., J.~A. {Burns}, V.~{Carruba}, P.~D. {Nicholson},\ and R.~A.
  {Jacobson} 2002.
\newblock {New Secular Resonances Involving the Irregular Satellies of Saturn}.
\newblock {\em AAS/Division of Dynamical Astronomy Meeting\/}~{\bf 33}.

\bibitem[{Dacosta}(1992){Dacosta}]{Dacosta.1992}
{Dacosta}, G.~S. 1992.
\newblock {Basic Photometry Techniques}.
\newblock In {\em ASP Conf. Ser. 23: Astronomical CCD Observing and Reduction
  Techniques}, pp.\  90--104.

\bibitem[{Dahlgren} {\em et~al.}(1997){Dahlgren}, {Lagerkvist}, {Fitzsimmons},
  {Williams}, and {Gordon}]{Dahlgren.1997}
{Dahlgren}, M., C.-I. {Lagerkvist}, A.~{Fitzsimmons}, I.~P. {Williams},\ and
  M.~{Gordon} 1997.
\newblock {A study of Hilda asteroids. II. Compositional implications from
  optical spectroscopy.}
\newblock {\em \aap\/}~{\bf 323}, 606--619.

\bibitem[{Davis} {\em et~al.}(1989){Davis}, {Weidenschilling}, {Farinella},
  {Paolicchi}, and {Binzel}]{Davis.1989}
{Davis}, D.~R., S.~J. {Weidenschilling}, P.~{Farinella}, P.~{Paolicchi},\ and
  R.~P. {Binzel} 1989.
\newblock {Asteroid collisional history - Effects on sizes and spins}.
\newblock In {\em Asteroids II}, pp.\  805--826.

\bibitem[{Degewij}(1980){Degewij}]{Degewij.1980a}
{Degewij}, J. 1980.
\newblock {Spectroscopy of faint asteroids, satellites, and comets}.
\newblock {\em \aj\/}~{\bf 85}, 1403--1412.

\bibitem[{Degewij} {\em et~al.}(1980){Degewij}, {Cruikshank}, and
  {Hartmann}]{Degewij.1980c}
{Degewij}, J., D.~P. {Cruikshank},\ and W.~K. {Hartmann} 1980.
\newblock {Near-infrared colorimetry of J6 Himalia and S9 Phoebe - A summary of
  0.3- to 2.2-micron reflectances}.
\newblock {\em Icarus\/}~{\bf 44}, 541--547.

\bibitem[{Degewij} {\em et~al.}(1980){Degewij}, {Zellner}, and
  {Andersson}]{Degewij.1980b}
{Degewij}, J., B.~{Zellner},\ and L.~E. {Andersson} 1980.
\newblock {Photometric properties of outer planetary satellites}.
\newblock {\em Icarus\/}~{\bf 44}, 520--540.

\bibitem[{Fujiwara} {\em et~al.}(1989){Fujiwara}, {Cerroni}, {Davis}, {Ryan},
  and {di Martino}]{Fujiwara.1989}
{Fujiwara}, A., P.~{Cerroni}, D.~{Davis}, E.~{Ryan},\ and M.~{di Martino} 1989.
\newblock {Experiments and scaling laws for catastrophic collisions}.
\newblock In {\em Asteroids II}, pp.\  240--265.

\bibitem[{Gladman} and {Boehnhardt}(2000){Gladman} and
  {Boehnhardt}]{Gladman.2000c}
{Gladman}, B.,\ and H.~{Boehnhardt} 2000.
\newblock {S/1999 J 1}.
\newblock {\em \iaucirc\/}~{\bf 7472}, 2.

\bibitem[{Gladman} {\em et~al.}(2000){Gladman}, {Kavelaars}, {Holman}, {Petit},
  {Scholl}, {Nicholson}, and {Burns}]{Gladman.2000a}
{Gladman}, B., J.~{Kavelaars}, M.~{Holman}, J.-M. {Petit}, H.~{Scholl},
  P.~{Nicholson},\ and J.~A. {Burns} 2000.
\newblock {NOTE: The Discovery of Uranus XIX, XX, and XXI}.
\newblock {\em Icarus\/}~{\bf 147}, 320--324.

\bibitem[{Gladman} {\em et~al.}(2001){Gladman}, {Kavelaars}, {Holman},
  {Nicholson}, {Burns}, {Hergenrother}, {Petit}, {Marsden}, {Jacobson}, {Gray},
  and {Grav}]{Gladman.2001}
{Gladman}, B., J.~J. {Kavelaars}, M.~{Holman}, P.~D. {Nicholson}, J.~A.
  {Burns}, C.~W. {Hergenrother}, J.~{Petit}, B.~G. {Marsden}, R.~{Jacobson},
  W.~{Gray},\ and T.~{Grav} 2001.
\newblock {Discovery of 12 satellites of Saturn exhibiting orbital clustering}.
\newblock {\em Nature\/}~{\bf 412}, 163--166.

\bibitem[{Gladman} {\em et~al.}(1998){Gladman}, {Nicholson}, {Burns},
  {Kavelaars}, {Marsden}, {Williams}, and {Offutt}]{Gladman.1998}
{Gladman}, B.~J., P.~D. {Nicholson}, J.~A. {Burns}, J.~J. {Kavelaars}, B.~G.
  {Marsden}, G.~V. {Williams},\ and W.~B. {Offutt} 1998.
\newblock {Discovery of two distant irregular moons of Uranus}.
\newblock {\em Nature\/}~{\bf 392}, 897--899.

\bibitem[{Green}(2002){Green}]{Green.2002}
{Green}, D.~W.~E. 2002.
\newblock {Satellites of Jupiter}.
\newblock {\em \iaucirc\/}~{\bf 7998}, 2.

\bibitem[{Greenberg} {\em et~al.}(1978){Greenberg}, {Hartmann}, {Chapman}, and
  {Wacker}]{Greenberg.1978}
{Greenberg}, R., W.~K. {Hartmann}, C.~R. {Chapman},\ and J.~F. {Wacker} 1978.
\newblock {Planetesimals to planets - Numerical simulation of collisional
  evolution}.
\newblock {\em Icarus\/}~{\bf 35}, 1--26.

\bibitem[{Hartmann} {\em et~al.}(1982){Hartmann}, {Cruikshank}, and
  {Degewij}]{Hartmann.1982}
{Hartmann}, W.~K., D.~P. {Cruikshank},\ and J.~{Degewij} 1982.
\newblock {Remote comets and related bodies - VJHK colorimetry and surface
  materials}.
\newblock {\em Icarus\/}~{\bf 52}, 377--408.

\bibitem[{Heppenheimer}(1975){Heppenheimer}]{Heppenheimer.1975}
{Heppenheimer}, T.~A. 1975.
\newblock {On the presumed capture origin of Jupiter's outer satellites}.
\newblock {\em Icarus\/}~{\bf 24}, 172--180.

\bibitem[{Heppenheimer} and {Porco}(1977){Heppenheimer} and
  {Porco}]{Heppenheimer.1977}
{Heppenheimer}, T.~A.,\ and C.~{Porco} 1977.
\newblock {New contributions to the problem of capture}.
\newblock {\em Icarus\/}~{\bf 30}, 385--401.

\bibitem[{Holman} {\em et~al.}(2002){Holman}, {Kavelaars}, {Milisavljevic},
  {Gladman}, {Nicholson}, {Dumas}, {Petit}, {Marsden}, {Rousselot}, {Mousis},
  and {Grav}]{Holman.2002}
{Holman}, M., J.~{Kavelaars}, D.~{Milisavljevic}, B.~{Gladman}, P.~{Nicholson},
  C.~{Dumas}, J.-M. {Petit}, B.~G. {Marsden}, P.~{Rousselot}, O.~{Mousis},\ and
  T.~{Grav} 2002.
\newblock {S/2001 U 1}.
\newblock {\em \iaucirc\/}~{\bf 7980}, 1.

\bibitem[{Howell}(1989){Howell}]{Howell.1989}
{Howell}, S.~B. 1989.
\newblock {Two-dimensional aperture photometry - Signal-to-noise ratio of
  point-source observations and optimal data-extraction techniques}.
\newblock {\em \pasp\/}~{\bf 101}, 616--622.

\bibitem[{Jarvis} {\em et~al.}(2000){Jarvis}, {Vilas}, {Larson}, and
  {Gaffey}]{Jarvis.2000a}
{Jarvis}, K.~S., F.~{Vilas}, S.~M. {Larson},\ and M.~J. {Gaffey} 2000.
\newblock {JVI Himalia: New compositional evidence and interpretations for the
  origin of Jupiter's small satellites}.
\newblock {\em Icarus\/}~{\bf 145}, 445--453.

\bibitem[{Kuiper}(1956){Kuiper}]{Kuiper.1956}
{Kuiper}, G.~P. 1956.
\newblock {On the origin of the satellites and the Trojans}.
\newblock {\em Vistas in Astronomy\/}~{\bf 2}, 1631--1666.

\bibitem[{Landolt}(1992){Landolt}]{Landolt.1992}
{Landolt}, A.~U. 1992.
\newblock {Broadband UBVRI photometry of the Baldwin-Stone Southern Hemisphere
  spectrophotometric standards}.
\newblock {\em \aj\/}~{\bf 104}, 372--376.

\bibitem[{Lassell}(1846){Lassell}]{Lassell.1846}
{Lassell}, W. 1846.
\newblock {Discovery of supposed ring and satellite of Neptune }.
\newblock {\em \mnras\/}~{\bf 7}, 157.

\bibitem[{Luu}(1991){Luu}]{Luu.1991}
{Luu}, J. 1991.
\newblock {CCD photometry and spectroscopy of the outer Jovian satellites}.
\newblock {\em \aj\/}~{\bf 102}, 1213--1225.

\bibitem[{Maris} {\em et~al.}(2001){Maris}, {Carraro}, {Cremonese}, and
  {Fulle}]{Maris.2001}
{Maris}, M., G.~{Carraro}, G.~{Cremonese},\ and M.~{Fulle} 2001.
\newblock {Multicolor Photometry of the Uranus Irregular Satellites Sycorax and
  Caliban}.
\newblock {\em \aj\/}~{\bf 121}, 2800--2803.

\bibitem[{McLeod} {\em et~al.}(2000){McLeod}, {Conroy}, {Gauron}, {Geary}, and
  {Ordway}]{McLeod.2000}
{McLeod}, B.~A., M.~{Conroy}, T.~M. {Gauron}, J.~C. {Geary},\ and M.~P.
  {Ordway} 2000.
\newblock {Megacam: A Wide-field Imager for the MMT Observatory}.
\newblock In {\em Further Developments in Scientific Optical Imaging}, pp.\
  ~11.

\bibitem[{Pollack} {\em et~al.}(1979){Pollack}, {Burns}, and
  {Tauber}]{Pollack.1979}
{Pollack}, J.~B., J.~A. {Burns},\ and M.~E. {Tauber} 1979.
\newblock {Gas drag in primordial circumplanetary envelopes - A mechanism for
  satellite capture}.
\newblock {\em Icarus\/}~{\bf 37}, 587--611.

\bibitem[{Rettig} {\em et~al.}(2001){Rettig}, {Walsh}, and
  {Consolmagno}]{Rettig.2001}
{Rettig}, T.~W., K.~{Walsh},\ and G.~{Consolmagno} 2001.
\newblock {Implied Evolutionary Differences of the Jovian Irregular Satellites
  from a BVR Color Survey}.
\newblock {\em Icarus\/}~{\bf 154}, 313--320.

\bibitem[{Romon} {\em et~al.}(2001){Romon}, {de Bergh}, {Barucci},
  {Doressoundiram}, {Cuby}, {Le Bras}, {Dout{\' e}}, and {Schmitt}]{Romon.2001}
{Romon}, J., C.~{de Bergh}, M.~A. {Barucci}, A.~{Doressoundiram}, J.-G. {Cuby},
  A.~{Le Bras}, S.~{Dout{\' e}},\ and B.~{Schmitt} 2001.
\newblock {Photometric and spectroscopic observations of Sycorax, satellite of
  Uranus}.
\newblock {\em \aap\/}~{\bf 376}, 310--315.

\bibitem[{Roy}(1988){Roy}]{Roy.1988}
{Roy}, A.~E. 1988.
\newblock {\em {Orbital motion}}.
\newblock Bristol, England ; Philadelphia : A.~Hilger, 1988.~3rd ed.

\bibitem[{Saha} and {Tremaine}(1993){Saha} and {Tremaine}]{Saha.1993}
{Saha}, P.,\ and S.~{Tremaine} 1993.
\newblock {The orbits of the retrograde Jovian satellites}.
\newblock {\em Icarus\/}~{\bf 106}, 549--562.

\bibitem[{Sawyer}(1991){Sawyer}]{Sawyer.1991}
{Sawyer}, S.~R. 1991.
\newblock {\em {A High-Resolution CCD Spectroscopic Survey of Low-Albedo Main
  Belt Asteroids.}}
\newblock Ph.\ D. thesis.

\bibitem[{Schaefer} and {Schaefer}(2000){Schaefer} and
  {Schaefer}]{Schaefer.2000}
{Schaefer}, B.~E.,\ and M.~W. {Schaefer} 2000.
\newblock {Nereid Has Complex Large-Amplitude Photometric Variability}.
\newblock {\em Icarus\/}~{\bf 146}, 541--555.

\bibitem[{Scotti} {\em et~al.}(2000){Scotti}, {Spahr}, {McMillan}, {Larsen},
  {Montani}, {Gleason}, {Gehrels}, {Marsden}, and {Williams}]{Scotti.2000}
{Scotti}, J.~V., T.~B. {Spahr}, R.~S. {McMillan}, J.~A. {Larsen}, J.~{Montani},
  A.~E. {Gleason}, T.~{Gehrels}, B.~G. {Marsden},\ and G.~V. {Williams} 2000.
\newblock {S/1999 J 1}.
\newblock {\em \iaucirc\/}~{\bf 7460}, 1.

\bibitem[{Sheppard} {\em et~al.}(2000){Sheppard}, {Jewitt}, {Fernandez},
  {Magnier}, {Marsden}, {Holman}, {Kowal}, {Roemer}, and
  {Williams}]{Sheppard.2000}
{Sheppard}, S.~S., D.~C. {Jewitt}, Y.~{Fernandez}, G.~{Magnier}, B.~G.
  {Marsden}, M.~{Holman}, C.~T. {Kowal}, E.~{Roemer},\ and G.~V. {Williams}
  2000.
\newblock {S/1975 J 1 = S/2000 J 1}.
\newblock {\em \iaucirc\/}~{\bf 7525}, 1.

\bibitem[{Sheppard} {\em et~al.}(2001){Sheppard}, {Jewitt}, {Fernandez},
  {Magnier}, {Marsden}, {Dahm}, and {Evans}]{Sheppard.2001}
{Sheppard}, S.~S., D.~C. {Jewitt}, Y.~R. {Fernandez}, G.~{Magnier}, B.~G.
  {Marsden}, S.~{Dahm},\ and A.~{Evans} 2001.
\newblock {Satellites of Jupiter}.
\newblock {\em \iaucirc\/}~{\bf 7555}, 1.

\bibitem[{Sheppard} {\em et~al.}(2002){Sheppard}, {Jewitt}, {Kleyna},
  {Marsden}, and {Jacobson}]{Sheppard.2002}
{Sheppard}, S.~S., D.~C. {Jewitt}, J.~{Kleyna}, B.~G. {Marsden},\ and
  R.~{Jacobson} 2002.
\newblock {Satellites of Jupiter}.
\newblock {\em \iaucirc\/}~{\bf 7900}, 1.

\bibitem[{Simonelli} {\em et~al.}(1999){Simonelli}, {Kay}, {Adinolfi},
  {Veverka}, {Thomas}, and {Helfenstein}]{Simonelli.1999}
{Simonelli}, D.~P., J.~{Kay}, D.~{Adinolfi}, J.~{Veverka}, P.~C. {Thomas},\ and
  P.~{Helfenstein} 1999.
\newblock {Phoebe: Albedo Map and Photometric Properties}.
\newblock {\em Icarus\/}~{\bf 138}, 249--258.

\bibitem[{S{\o}rensen} and {N{\o}rregaard}(1996){S{\o}rensen} and
  {N{\o}rregaard}]{Sorensen.1996}
{S{\o}rensen}, N.~A.,\ and P.~{N{\o}rregaard} 1996.
\newblock {The ALFOSC camera, properties of the W11-3 CCD.}
\newblock Technical report, {Copenhagen University Observatory}.

\bibitem[{Stetson} {\em et~al.}(1990){Stetson}, {Davis}, and
  {Crabtree}]{Stetson.1990}
{Stetson}, P.~B., L.~E. {Davis},\ and D.~R. {Crabtree} 1990.
\newblock {Future development of the DAOPHOT crowded-field photometry package}.
\newblock In {\em ASP Conf. Ser. 8: CCDs in astronomy}, pp.\  289--304.

\bibitem[{Sykes} {\em et~al.}(2000){Sykes}, {Nelson}, {Cutri}, {Kirkpatrick},
  {Hurt}, and {Skrutskie}]{Sykes.2000}
{Sykes}, M.~V., B.~{Nelson}, R.~M. {Cutri}, D.~J. {Kirkpatrick}, R.~{Hurt},\
  and M.~F. {Skrutskie} 2000.
\newblock {Near-Infrared Observations of the Outer Jovian Satellites}.
\newblock {\em Icarus\/}~{\bf 143}, 371--375.

\bibitem[{Tegler} and {Romanishin}(1998){Tegler} and {Romanishin}]{Tegler.1998}
{Tegler}, S.~C.,\ and W.~{Romanishin} 1998.
\newblock {Two distinct populations of Kuiper-belt objects}.
\newblock {\em \nat\/}~{\bf 392}, 49.

\bibitem[{Tegler} and {Romanishin}(2000){Tegler} and {Romanishin}]{Tegler.2000}
{Tegler}, S.~C.,\ and W.~{Romanishin} 2000.
\newblock {Extremely red Kuiper-belt objects in near-circular orbits beyond 40
  AU}.
\newblock {\em \nat\/}~{\bf 407}, 979--981.

\bibitem[{Tholen} and {Zellner}(1983){Tholen} and {Zellner}]{Tholen.1983}
{Tholen}, D.~J.,\ and B.~{Zellner} 1983.
\newblock {Eight-color photometry of Hyperion, Iapetus, and Phoebe}.
\newblock {\em Icarus\/}~{\bf 53}, 341--347.

\bibitem[{Tholen} and {Zellner}(1984){Tholen} and {Zellner}]{Tholen.1984}
{Tholen}, D.~J.,\ and B.~{Zellner} 1984.
\newblock {Multicolor photometry of outer Jovian satellites}.
\newblock {\em Icarus\/}~{\bf 58}, 246--253.

\bibitem[{Tody}(1986){Tody}]{Tody.1986}
{Tody}, D. 1986.
\newblock {The IRAF Data Reduction and Analysis System}.
\newblock In {\em Instrumentation in astronomy VI; Proceedings of the Meeting,
  Tucson, AZ, Mar. 4-8, 1986. Part 2 (A87-36376 15-35). Bellingham, WA, Society
  of Photo-Optical Instrumentation Engineers, 1986, p. 733.}, Volume 627, pp.\
  733.

\bibitem[{Tody}(1993){Tody}]{Tody.1993}
{Tody}, D. 1993.
\newblock {IRAF in the Nineties}.
\newblock In {\em ASP Conf. Ser. 52: Astronomical Data Analysis Software and
  Systems II}, Volume~2, pp.\  173.

\end{thebibliography}

\newpage

\section*{Table Caption}

Table \ref{tab:obs} Here are shown the observational circumstances for each
object. OPT is the Observer-Primary-Target angle.

Table \ref{tab:res} Here are shown the photometric results 
of the observations. The $m_v(1,1,0)$ gives the absolute
magnitude normalized to unit heliocentric and geocentric
distances, and zero phase angle. $1\sigma$ errors are given.

\section*{Figure Caption}

\begin{figure}[p]
  \begin{center}
    \leavevmode
    \includegraphics[width=6in, height=6in]{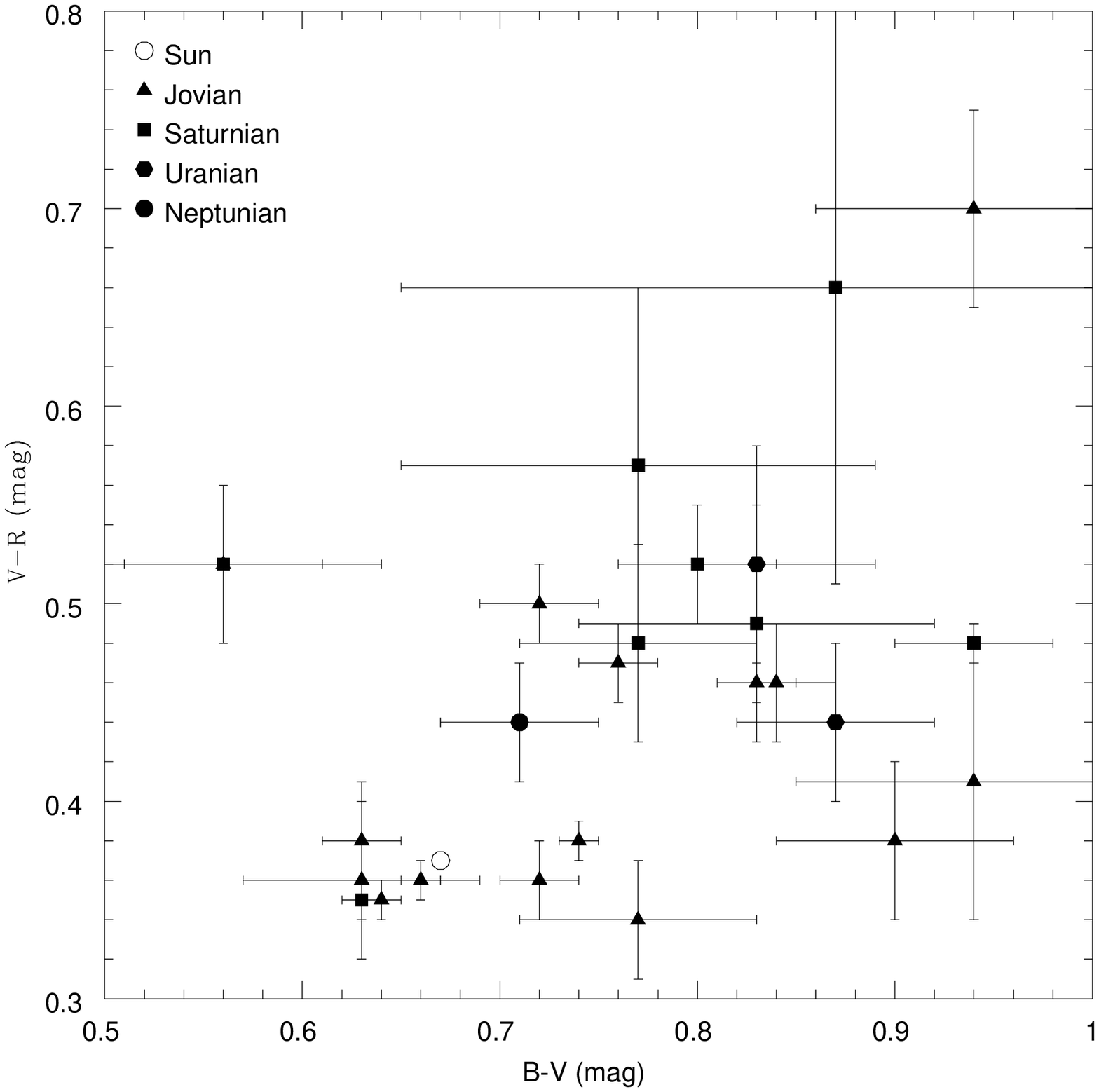}
  \end{center}
  \caption[ND]{Grav et al., Photometric Survey of Irregular Satellites}
  \label{fig:bvvr}
\end{figure}

Figure \ref{fig:bvvr}. The $B-V$ vs. $V-I$ plot of all available 
photometry of irregular satellites. Photometry of J~VI~Himalia,
J~VII~Elara, J~XIII~Leda is from \citet{Rettig.2001} and the two
Uranian U~XVI~Caliban and U~XVII~Sycorax can be found in 
\citet{Maris.2001,Romon.2001}. The colos of N~II~Nereid can be
found in \citet{Schaefer.2000}. Note that J~XX~Taygete and S/2000~S~1
have identical $B-V$ and $V-I$ colors. 

\begin{figure}[p]
  \begin{center}
    \leavevmode
    \includegraphics[width=6in, height=6in]{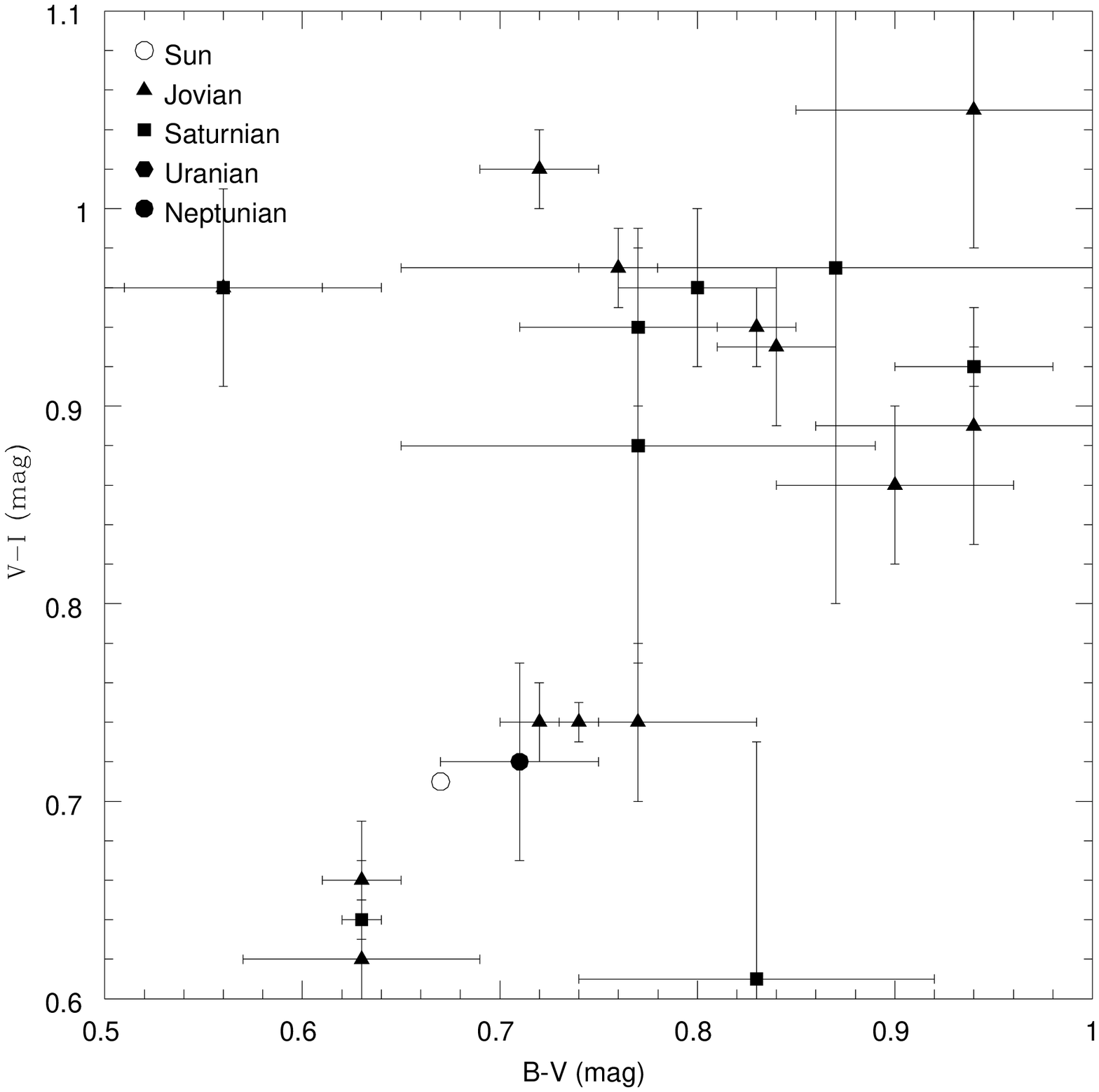}
  \end{center}
  \caption[ND]{Grav et al., Photometric Survey of Irregular Satellites}
  \label{fig:bvvi}
\end{figure}

Figure \ref{fig:bvvi}. The same as Fig. \ref{fig:bvvr}, but showing the 
$B-V$ vs. $V-I$ colors.

\begin{figure}[p]
  \begin{center}
    \leavevmode
    \includegraphics[width=6in, height=6in]{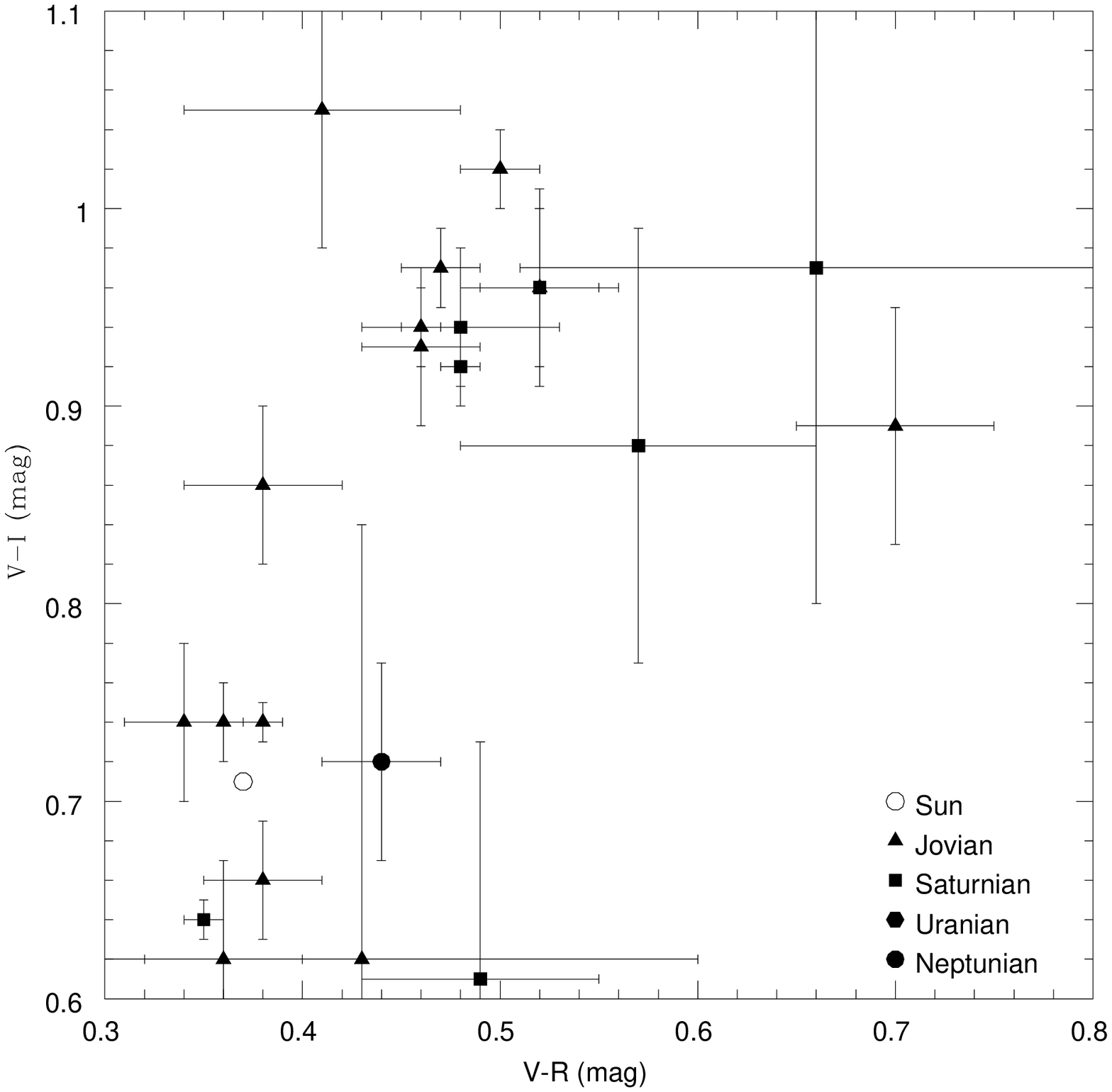}
  \end{center}
  \caption[ND]{Grav et al., Photometric Survey of Irregular Satellites}
  \label{fig:vrvi}
\end{figure}

Figure \ref{fig:vrvi}. The same as Fig. \ref{fig:bvvr}, but showing the 
$V-R$ vs. $V-I$ colots.

\begin{figure}[p]
  \begin{center}
    \leavevmode
    \includegraphics[width=6in, height=6in]{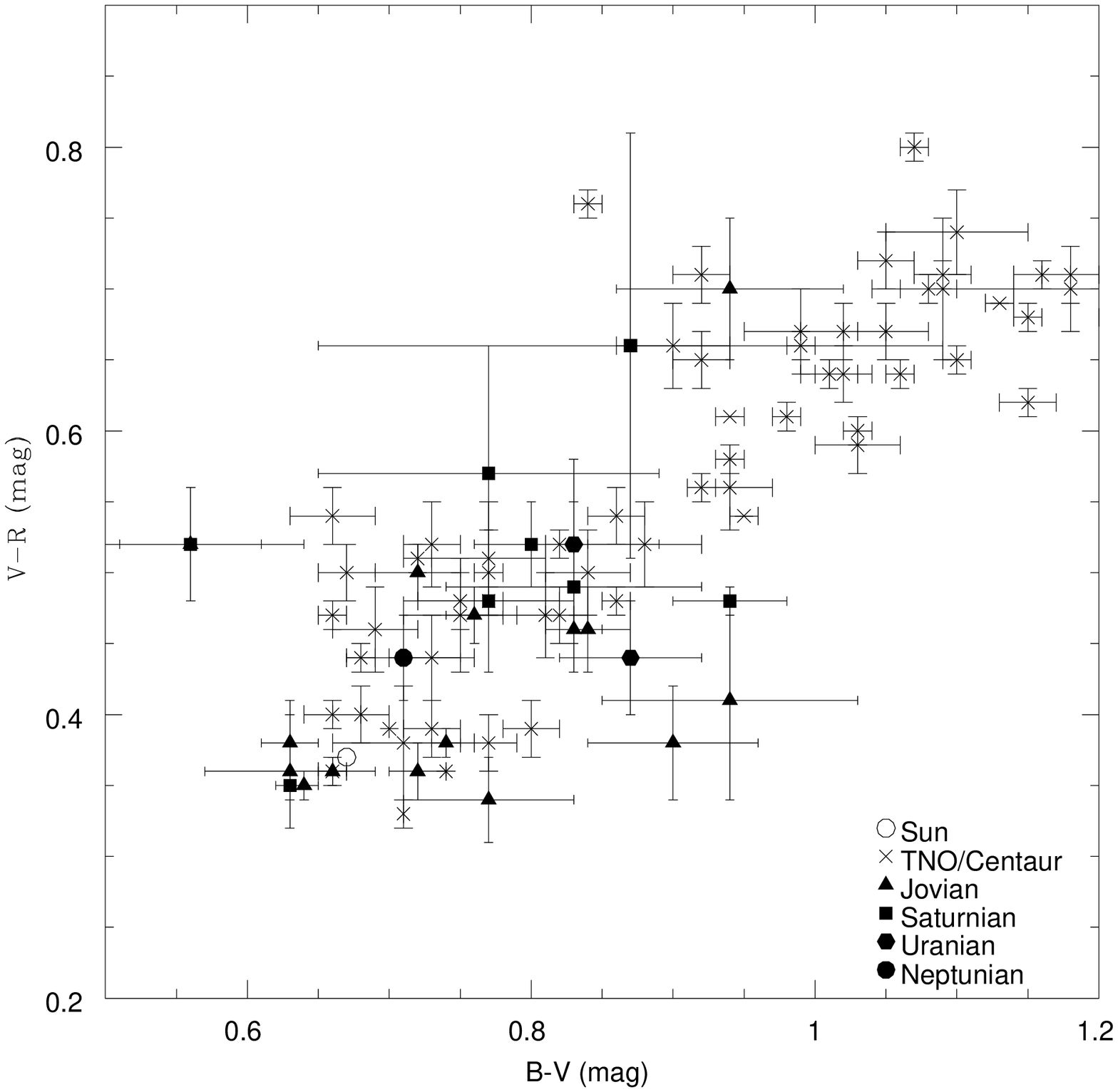}
  \end{center}
  \caption[ND]{Grav et al., Photometric Survey of Irregular Satellites}
  \label{fig:bvvrtno}
\end{figure}

Figure \ref{fig:bvvrtno}. The same as Fig. \ref{fig:bvvr} but also
containing the known Centaurs and TNOs with photometry better than
$0.05$ magnitude. It is seen that the colors of the irregular satellites
are similar to the {\it grey} class of the Centaurs and TNOs 
\citep{Tegler.1998,Tegler.2000}. 

\begin{figure}[p]
  \begin{center}
    \leavevmode
    \includegraphics[width=6in, height=6in]{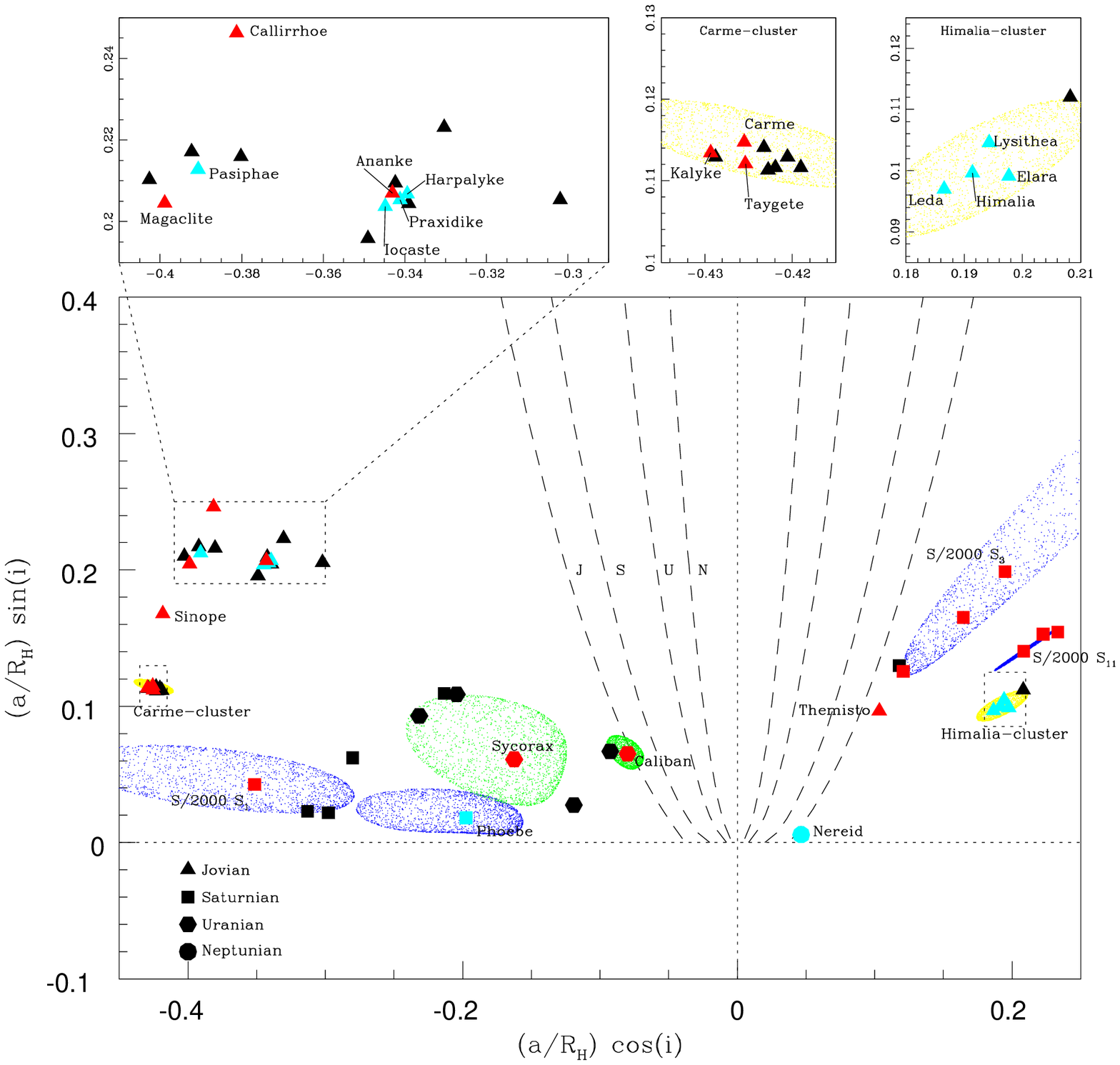}
  \end{center}
  \caption[ND]{Grav et al., Photometric Survey of Irregular Satellites}
  \label{fig:polar}
\end{figure}

Figure \ref{fig:polar}. Here is shown the mean motion semimajor
axis and inclination of the irregular satellites plotted in a polar
plot normalized to the Hill-sphere radius of the satellite's parent
planet. The irregular satellites for which colors have been determined
is plotted in color depending on which color class we put it in, {\it blue} 
for {\bf grey} and {\it red} for {\bf light red}. The colored ellipses give the
area of a-i phase space were each cluster could disperse to given an
catastrophic event which gives the fragments speeds of $135$m/s for
J~VI~Himalia, $30$m/s for J XI Carme, $50$m/s for the Saturnian
$34^\circ$ cluster, $350$m/s for the Saturnian $45^\circ$ cluster,
$200$m/s for the S~IX~Phoebe, $120$m/s for the Saturnian $174^\circ$
cluster, $75$m/s and fragmentation at $M=180^\circ$ for U~XVI~Caliban,
and $120$m/s and fragmentation at $M=80^\circ$ for U~XVII~Sycorax.
The region within which the the Kozai mechanism drives orbits to 
eccentricities high enough for the objects to penetrate into the 
regular satellites is indicated by the dashed lines for each planet
\citep{Gladman.2001}.  

\newpage

\setlength{\parindent}{0pt}
\begin{table}[p]
\begin{center}
\begin{tabular}{lrlrrrrr}
 Object     & Telescope     &  UT          &  R   & $\Delta$ & $\alpha$     & OPT \\
            &               & Date         & (AU) & (AU)     & (${}^\circ$) & (${}^\circ$) \\
 \hline
 Pasiphae   & Magellan-6.5m & 2002 Feb. 12.& 5.10 & 4.37     &  8.01        &  47.5 \\
 Sinope     & NOT-2.56m     & 2002 Jan. 14.& 5.30 & 4.33     &  2.35        & 124.3 \\
 Lysithea   & Magellan-6.5m & 2002 Feb. 12.& 5.18 & 4.46     &  8.14        &  89.3 \\
 Carme      & NOT-2.56m     & 2002 Mar. 14.& 5.18 & 4.86     & 10.75        &  73.0 \\ 
 Ananke     & NOT-2.56m     & 2002 Mar. 14.& 5.11 & 4.82     & 10.99        &  33.2 \\
 Callirrhoe & Magellan-6.5m & 2002 Feb. 09.& 5.34 & 4.58     &  7.23        & 153.9 \\
 Themisto   & Magellan-6.5m & 2002 Feb. 12.& 5.16 & 4.44     &  8.10        &  64.6 \\
 Kalyke     & NOT-2.56m     & 2002 Jan. 14.& 5.04 & 4.09     &  3.16        &  41.8 \\
            & MMT-6m        & 2002 Mar. 13.& 5.05 & 4.73     & 11.06        &  22.5 \\
 Iocaste    & Magellan-6.5m & 2002 Feb. 09.& 5.06 & 4.31     &  7.89        &  35.0 \\ 
 Harpalyke  & NOT-2.56m     & 2002 Jan. 14.& 5.15 & 4.20     &  3.11        &  80.4 \\
 Praxidike  & Magellan-6.5m & 2002 Feb. 12.& 5.06 & 4.33     &  8.11        &  30.1 \\
 Magaclite  & NOT-2.56m     & 2002 Jan. 14.& 5.17 & 4.21     &  2.27        &  85.2 \\
 Taygete    & Magellan-6.5m & 2002 Feb. 09.& 5.02 & 4.27     &  7.94        &  21.4 \\
\\	   
 Phoebe     & NOT-2.56m     & 2001 Feb. 18.& 9.08 & 9.13     &  6.21        &  76.4 \\
 S/2000 S1  & Magellan-6.5m & 2002 Feb. 09.& 9.20 & 8.86     &  5.86        & 158.6 \\
 S/2000 S2  & NOT-2.56m     & 2002 Jan. 13.& 9.06 & 8.33     &  4.36        &  85.9 \\
 S/2000 S3  & NOT-2.56m     & 2002 Jan. 13.& 8.93 & 8.22     &  4.49        &  39.1 \\
 S/2000 S4  & NOT-2.56m     & 2002 Jan. 14.& 8.95 & 8.22     &  4.40        &  50.3 \\
 S/2000 S5  & NOT-2.56m     & 2002 Mar. 14.& 9.03 & 9.23     &  6.12        &  62.8 \\
 S/2000 S10 & MMT-6m        & 2002 Mar. 13.& 8.93 & 9.11     &  6.21        &  39.7 \\
 S/2000 S11 & NOT-2.56m     & 2001 Feb. 18.& 9.18 & 9.23     &  6.15        & 120.8 \\
            & NOT-2.56m     & 2002 Jan. 14.& 9.04 & 8.31     &  4.39        &  66.0 \\
\end{tabular}
   \caption[ND] {Grav et al., Photometric Survey of Irregular Satellites}
   \label{tab:obs}
\end{center}
\end{table}

\begin{table}[p]
\begin{center}
\begin{tabular}{lrccccc}
 Object   & Inclination & $m_V$ (1,1,0)        
    & V & B-V & V-R & V-I \\
\hline
Lysithea     & $28.3^{\circ}$  & $11.09\pm0.02$
    & $18.48\pm0.02$ & $0.72\pm0.02$ & $0.36\pm0.02$ & $0.74\pm0.02$ \\
Themisto     & $43.1^{\circ}$  & $12.94\pm0.01$
    & $20.31\pm0.01$ & $0.83\pm0.02$ & $0.46\pm0.01$ & $0.94\pm0.02$ \\
Callirrhoe   & $147.1^{\circ}$ & $13.92\pm0.02$
    & $21.39\pm0.02$ & $0.72\pm0.03$ & $0.50\pm0.02$ & $1.02\pm0.02$ \\
Harpalyke    & $148.7^{\circ}$ & $16.03\pm0.12$
    & $23.02\pm0.12$ &               & $0.43\pm0.17$ & $0.62\pm0.22$ \\
Praxidike    & $148.7^{\circ}$ & $15.24\pm0.03$
    & $22.51\pm0.03$ & $0.77\pm0.06$ & $0.34\pm0.03$ & $0.74\pm0.04$ \\ 
Ananke       & $148.9^{\circ}$ & $11.87\pm0.03$
    & $19.51\pm0.03$ & $0.90\pm0.06$ & $0.38\pm0.04$ & $0.86\pm0.04$ \\
Iocaste      & $149.7^{\circ}$ & $15.27\pm0.03$
    & $22.52\pm0.03$ & $0.63\pm0.06$ & $0.36\pm0.04$ & $0.62\pm0.05$ \\
Pasiphae     & $151.4^{\circ}$ & $ 9.92\pm0.01$
    & $17.22\pm0.01$ & $0.74\pm0.01$ & $0.38\pm0.01$ & $0.74\pm0.01$ \\
Magaclite    & $152.7^{\circ}$ & $15.12\pm0.05$
    & $22.07\pm0.05$ & $0.94\pm0.09$ & $0.41\pm0.07$ & $1.05\pm0.07$ \\
Sinope       & $158.1^{\circ}$ & $11.56\pm0.02$
    & $18.63\pm0.02$ & $0.84\pm0.03$ & $0.46\pm0.03$ & $0.93\pm0.04$ \\ 
Carme        & $164.9^{\circ}$ & $10.91\pm0.02$
    & $18.59\pm0.02$ & $0.76\pm0.02$ & $0.47\pm0.02$ & $0.97\pm0.02$ \\
Kalyke       & $165.2^{\circ}$ & $15.42\pm0.07$
    & $22.31\pm0.07$ &               & $0.72\pm0.09$ & $0.96\pm0.16$ \\
             &                      & $15.28\pm0.04$
    & $22.86\pm0.04$ & $0.94\pm0.08$ & $0.70\pm0.05$ & $0.88\pm0.06$ \\
Taygete      & $165.2^{\circ}$ & $15.63\pm0.04$
    & $22.85\pm0.04$ & $0.56\pm0.08$ & $0.52\pm0.04$ & $0.96\pm0.05$ \\
\\
S/2000 S4  & $33.5^{\circ}$    & $12.56\pm0.06$           
    & $22.28\pm0.06$ & $0.77\pm0.12$ & $0.57\pm0.09$ & $0.88\pm0.11$ \\
S/2000 S11 & $34.0^{\circ}$    & $10.88\pm0.03$    
    & $21.00\pm0.03$ & $0.89\pm0.07$ & $0.50\pm0.05$ & $0.91\pm0.05$ \\
             &                         & $11.20\pm0.03$
    & $20.97\pm0.03$ & $0.98\pm0.07$ & $0.47\pm0.04$ & $0.92\pm0.04$ \\
S/2000 S10 & $34.5^{\circ}$    & $13.90\pm0.06$
    & $23.90\pm0.05$ & $0.83\pm0.09$ & $0.49\pm0.06$ & $0.61\pm0.12$ \\
S/2000 S2  & $45.1^{\circ}$   & $11.83\pm0.03$
    & $21.61\pm0.03$ & $0.77\pm0.06$ & $0.48\pm0.05$ & $0.94\pm0.04$ \\
S/2000 S3  & $45.6^{\circ}$    & $10.69\pm0.02$
    & $20.38\pm0.02$ & $0.80\pm0.04$ & $0.52\pm0.03$ & $0.96\pm0.04$ \\
S/2000 S5  & $46.2^{\circ}$    & $12.65\pm0.12$             
    & $22.73\pm0.12$ & $0.87\pm0.22$ & $0.66\pm0.15$ & $0.97\pm0.17$ \\
S/2000 S1  & $173.1^{\circ}$   & $12.39\pm0.03$
    & $22.41\pm0.03$ & $0.56\pm0.05$ & $0.52\pm0.04$ & $0.96\pm0.05$ \\
Phoebe     & $174.8^{\circ}$   & $6.63\pm0.01$
    & $16.71\pm0.01$ & $0.63\pm0.01$ & $0.35\pm0.01$ & $0.64\pm0.01$ \\

\end{tabular}
   \caption[ND] {Grav et al., Photometric Survey of Irregular Satellites}
   \label{tab:res} 
\end{center}
\end{table}


\end{document}